\documentclass[prl,onecolumn, superscriptaddress,amsmath,amssymb, aps,10pt]{revtex4-2}

\usepackage{graphicx,bm,braket,xcolor}

\usepackage{hyperref}
\DeclareUnicodeCharacter{2212}{-}
\usepackage{float}
\usepackage{comment}
\usepackage{booktabs}
\usepackage{tabularx}

\begin{document}

%\title{All-optical modulation and long-lived tuning of diamond photonic cavities}
\title{Optically-Induced Modulation and Programming of Diamond Photonic Cavities}

\author{Yuchun Zhu}
\author{Amirali Arabmoheghi}
\affiliation{Institute of Physics, Swiss Federal Institute of Technology Lausanne (EPFL), CH-1015 Lausanne, Switzerland} 
\affiliation{Center of Quantum Science and Engineering, Swiss Federal Institute of Technology Lausanne (EPFL), CH-1015 Lausanne, Switzerland
}%
\author{Nicolas Le Thomas}
\affiliation{Photonics Research Group, Department of Information Technology (INTEC), Ghent University - IMEC, Ghent, Belgium}
\author{Niels Quack}
\affiliation{Institute of Microelectronics, University of Stuttgart, 70569 Stuttgart, Germany} 
\author{Valentin Goblot}
\email{valentin.goblot@epfl.ch}
\author{Christophe Galland}
\email{chris.galland@epfl.ch}
\affiliation{Institute of Physics, Swiss Federal Institute of Technology Lausanne (EPFL), CH-1015 Lausanne, Switzerland} 
\affiliation{Center of Quantum Science and Engineering, Swiss Federal Institute of Technology Lausanne (EPFL), CH-1015 Lausanne, Switzerland
}%
\date{\today}

%%%%%%%%%%%%%%%%%%%%%%%%%%%%%%%%%%%%%%%%%%%%%%%%%%%%%%%%%%%%%%%%%%%%%%%%%%%%%%%%%%%%%%%%%%%%%%%%%%%%%%%%%%%%%%%%%%%%

\begin{abstract}
Single-crystal diamond combines exceptional optical, thermal, and mechanical properties while hosting optically addressable and spin-coherent colour centres, making it a promising platform for integrated quantum and nonlinear photonics. However, practical post-fabrication mechanisms for dynamically controlling and tuning monolithic diamond photonic circuits remain limited. 
Here, we demonstrate all-optical modulation and long-lived tuning of suspended diamond Fabry--Perot nanobeam cavities containing nitrogen-vacancy centres. Under 532~nm illumination, the infrared (1000 to 1100~nm) cavity response is governed by two competing contributions: a well-understood thermo-optic red-shift, and a novel charge-mediated blue-shift that we attribute to the free-carrier plasma dispersion effect. 
With 10~mW of 20~kHz-modulated green light, we achieve an infrared modulation depth of approximately 45\% of the available reflection contrast. 
Green illumination also produces quasi-permanent photo-refractive resonance blue-shifts, with a maximum observed tuning of 3.15~nm (0.87~THz), without measurable degradation of the cavity linewidth or contrast. 
Achieving the same cavity shift in a lithium-niobate Pockels modulator would require 200 to 500~V across a \(5~\mu\mathrm{m}\) electrode gap.
These results establish a monolithic diamond nanophotonic platform combining agile and reversible modulation with long-lived optical reconfiguration, two key ingredients in photonic technologies.
\end{abstract}

\maketitle

%%%%%%%%%%%%%%%%%%%%%%%%%%%%%%%%%%%%%%%%%%%%%%%%%%%%%%%%%%%%%%%%%%%%%%%%%%%%%%%%%%%%%%%%
%%%%%%%%%%%%%%%%%%%%%%%%%%%%%%%%%%%%%%%%%%%%%%%%%%%%%%%%%%%%%%%%%%%%%%%%%%%%%%%%%%%%%%%%
\section{Introduction}
\label{sec:introduction}

The growing commercial availability of synthetic single-crystal diamond with controlled properties and dimensions~\cite{kiss2021diamondDiffractive}, together with advances in substrate thinning and thin-film processing~\cite{piracha2016scalable,guo2021tunable}, has enabled the development of increasingly complex diamond photonic integrated circuits~\cite{Sipahigil2016,Alison2021,Masuda2024,Ding2024,Chen2024}. Its large electronic bandgap of approximately \(5.5~\mathrm{eV}\) and nonpolar crystal lattice support optical transparency from the ultraviolet to the mid-infrared~\cite{lenzini2018diamondPlatform}. Bulk diamond also possesses exceptionally high thermal conductivity, although heat transport becomes strongly size-dependent in suspended micro- and nanostructures~\cite{goblot2026heatTransportDiamond}. Together with its high optical-damage threshold, these properties have enabled nonlinear photonic devices including on-chip Raman lasers and frequency-conversion structures~\cite{latawiec2015onchip,hausmann2014diamondNonlinear}.
Additionally, diamond hosts a broad range of optically active point defects whose electronic and spin states can be addressed using light. Among these, the negatively charged nitrogen-vacancy (NV) centre has become an important system for quantum information processing, nanoscale sensing and optically detected magnetic resonance~\cite{Doherty2013NV,Schirhagl2014Nanoscale}. Integrating such defects with waveguides and optical cavities can enhance their interaction with light and provide compact interfaces between optical, spin and microwave degrees of freedom~\cite{Faraon2012CouplingNV,Li2015CoherentSpin}. Recent cavity designs have further emphasised efficient transfer between strongly confined cavity modes and propagating waveguide or fibre modes, including corrugated photonic-crystal architectures developed for colour-centre interfacing~\cite{Bopp2024Sawfish,Bopp2025Dinosaur}. However, extending diamond photonics from passive structures towards actively controlled and reconfigurable optical circuits remains challenging.

In established photonic platforms, active control is commonly achieved through thermo-optics, Pockels effect, or free-carrier plasma dispersion. All-optical thermo-optic tuning has been demonstrated in a hybrid diamond--silicon microdisk resonator~\cite{Hill2020AllOpticalTuning}, but this approach is linked with high energy consumption and limited bandwidth, and can introduce thermal crosstalk in dense circuits. 
Pockels electro-optic modulators provide rapid and energy-efficient control, but pristine diamond is centrosymmetric and therefore does not possess a bulk Pockels effect. And even if NV-doped diamond has been reported to generate second-harmonic   \cite{Abulikemu2021SecondHarmonicDiamond,abulikemu2022temperature,Flagan2025OpticalSwitchingDiamond,barhum2026}, electro-optic modulation has not been reported to date.  
Mechanical, strain-mediated, and hybrid approaches provide alternative routes towards active diamond devices. Mechanically driven resonators and surface-acoustic-wave structures have enabled coherent control of NV and SiV spin transitions~\cite{barfuss2015strong,golter2016saw,maity2020acoustic}, while hybrid piezoelectric platforms have demonstrated gigahertz-frequency phonon delivery and strain control of waveguide-coupled colour centres~\cite{ding2024phononic,clark2024nanoelectromechanical}.

Photoinduced charge dynamics offer a distinct route to active photonics in defect-rich diamond.
Carrier-based modulation is the workhorse of silicon photonics~\cite{reed2005,Hsu2024SubVoltMOSCAPMicroring}, where thermally activated shallow dopants provide free carriers, whose concentration is modulated by an external electric field. Unfortunately, this approach does not directly translate to diamond and many other wide bandgap materials, where ambipolar doping with shallow impurities and low-resistance electrical carrier injection are considerably more challenging.
Yet, visible illumination can ionize deep impurities in diamond, modifying the charge states of NV centres, substitutional nitrogen, or other defects. Free carriers can be generated in the process. They may subsequently migrate and become trapped at bulk defects or surfaces. 
These processes have been observed through photoconductive measurements, spatial charge patterning, and charge-state conversion in bulk and near-surface diamond~\cite{Bourgeois2015PhotoelectricDetectionNV,Jayakumar2016OpticalPatterningTrappedCharge,Dhomkar2018ChargeDynamicsNV}. Optical modification of NV charge states has also been used to switch a defect-mediated effective second-order nonlinearity in a diamond microcavity~\cite{Flagan2025OpticalSwitchingDiamond}. In a nanophotonic structure, similar optical effects have been proposed to induce internal electric fields that modify the refractive index. Long-lived photoinduced resonance tuning has recently been observed in monolithic diamond nanocavities, including nonvolatile tuning and large spectral shifts under prolonged illumination~\cite{Itoi2025NonvolatilePhotorefractiveDiamond,Itoi2026LargePhotorefractiveShift}. 

Here, we realize optically induced tuning and reversible modulation of NV-rich suspended single-crystal diamond Fabry--Perot nanobeam cavities.  
Infrared (IR) laser light is coupled from one side of the nanobeam using a tapered optical fibre, while the control \(532~\mathrm{nm}\) light is focused onto the device from the top. As recently reported in similar structures \cite{Itoi2025NonvolatilePhotorefractiveDiamond}, repeated green illumination produces a long-lived shift of the cavity resonances towards shorter wavelengths.  
Importantly, this tuning occurs without measurable degradation of the cavity linewidth or reflection contrast. The temporal response is consistent with charge diffusion along the waveguide leading to the formation of a metastable local charge configuration. 
More surprisingly, time-resolved measurements reveal rich refractive index dynamics. In addition to the expected transient thermo-optic red-shift, we discover a competing and reversible blue-shift accompanied by an increase in absorption, which are both consistent with the free-carrier plasma dispersion effect, whose relative strength can be controlled by illumination geometry and intensity.
These results show that the same green excitation can access both persistent post-fabrication tuning and fast reversible modulation through a combination of thermo-optic and photo-carrier responses. Their coexistence provides a basis for reconfigurable and dynamically controlled diamond photonic circuits, while highlighting how the rich photophysics of bulk impurities and surface defects can impact light propagation in diamond photonic nanostructures. Our observations open new research directions toward active photonics circuits and optoelectronics with doped wide bandgap materials. 

%%%%%%%%%%%%%%%%%%%%%%%%%%%%%%%%%%%%%%%%%%%%%%%%%%%%%%%%%%%%%%%%%%%%%%%%%%%%%%%%%%%%%%%%
%%%%%%%%%%%%%%%%%%%%%%%%%%%%%%%%%%%%%%%%%%%%%%%%%%%%%%%%%%%%%%%%%%%%%%%%%%%%%%%%%%%%%%%%
\section{Results}

%%%%%%%%%%%%%%%%%%%%%%%%%%%%%%%%%%%%%%%%%%%%%%%%%%%%%%%%%%%%%%%%%%%%%%%%%%%%%%%%%%%%%%%%
\begin{figure}[t]
    \centering
    \includegraphics[width=\linewidth]{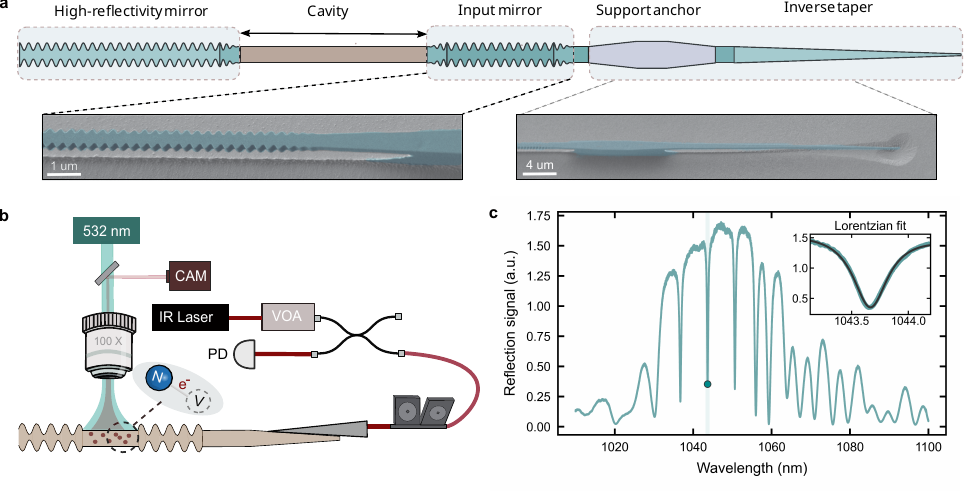}
    \caption{
    \textbf{Experimental scheme and representative cavity spectrum.}
    \textbf{a,} Schematic of a suspended diamond Fabry--Perot nanobeam cavity, comprising a high-reflectivity Bragg mirror (anchored to the bulk), a cavity, a lower-reflectivity input mirror, a support anchor, and an inverse taper. The nominal nanobeam cross-section is \(437~\mathrm{nm}\) wide and \(340~\mathrm{nm}\) thick for single-mode operation at infrared wavelengths. Insets are scanning electron micrographs of a fabricated device. The complete set of parameters is provided in Supplementary Table~\ref{tab:fp_design_parameters}.
    \textbf{b,} Experimental configuration for fibre-coupled IR reflection spectroscopy under local $532~\mathrm{nm}$ excitation. Tunable IR light is passed through a variable optical attenuator (VOA) and coupled into the device through an etched optical fibre taper. Paddles are used to control the polarization and excite the TE mode. Light reflected from the cavity goes through the beam splitter to a photodetector. The green beam is focused onto the cavity through a microscope objective, and a camera (CAM) is used to image the device and align the excitation spot.
    \textbf{c,} Representative experimental IR reflection spectrum of the fabricated cavity. The highlighted resonance near \(1043.7~\mathrm{nm}\) is enlarged in the inset together with a Lorentzian fit.
    }
    \label{fig:structure-setup}
\end{figure}
%%%%%%%%%%%%%%%%%%%%%%%%%%%%%%%%%%%%%%%%%%%%%%%%%%%%%%%%%%%%%%%%%%%%%%%%%%%%%%%%%%%%%%%%

%%%%%%%%%%%%%%%%%%%%%%%%%%%%%%%%%%%%%%%%%%%%%%%%%%%%%%%%%%%%%%%%%%%%%%%%%%%%%%%%%%%%%%%%
\subsection{Device fabrication and measurement}
\label{subsec:device_fabrication_measurement}

The cavity geometry was designed to obtain a longitudinal resonance near the $1042~\mathrm{nm}$ singlet transition of negatively charged NV centres, with the prospect of using the device for absorption-based spin readout ~\cite{Acosta2010Broadband,Jensen2014CavityEnhanced}. All results presented here, however, are reproducibly observed across resonances spanning 1000 to 1100~nm, only limited by our laser tuning range, excluding a key role of the singlet absorption. The devices were fabricated in single-crystal diamond with a manufacturer-specified NV concentration of $4~\mathrm{ppm}$ (Element Six)~\cite{ElementSix2021DNVB14}. 
The device geometry is illustrated in Fig.~\ref{fig:structure-setup}a. Each device consists of a suspended Fabry--Perot nanobeam cavity defined by two corrugated Bragg mirrors (tapered to reduce scattering losses). Two sets of cavity lengths -- \(20\) and \(50~\mu\mathrm{m}\) -- are discussed here. A high-reflectivity mirror terminates one side of the cavity, whereas a lower-reflectivity input mirror provides optical access from the fibre coupling side, consisting of a support anchor and an inverse taper \cite{burek2017fiber}.
After the device pattern is transferred into the diamond by electron-beam lithography and vertical etch, a quasi-isotropic oxygen-plasma undercut etch was used to suspend the nanobeams. Similar approaches have been used to fabricate suspended microdisks and rectangular nanobeam cavities out of bulk diamond \cite{khanaliloo2015microdisks,mouradian2017rectangular}. The resulting nanobeam is suspended approximately $1~\mu\mathrm{m}$ above the underlying diamond substrate. The scanning electron microscopy images in Fig.~\ref{fig:structure-setup}a show a Bragg mirror (left) and a support anchor with the inverse-taper coupling section (right). All design parameters are listed in Table~\ref{tab:fp_design_parameters}.

The optical measurement configuration is shown schematically in Fig.~\ref{fig:structure-setup}b. Light from a continuous-wave tunable IR laser (Toptica) is coupled into the suspended diamond nanobeam using an etched optical fibre taper aligned with the inverse taper of the diamond device. Reflection from the cavity is collected through the same fibre and directed to a photodetector through a beam-splitter.
A $532~\mathrm{nm}$ laser beam (Cobolt) was delivered from free space through a microscope objective with a numerical aperture of $0.5$ (Mitutoyo) and focused onto the tested device. In some measurements, a more homogeneous illumination of the entire cavity was achieved by introducing a cylindrical lens into the excitation path to reshape the initially near-circular spot into an elliptical one. For time-resolved measurements, the green laser was electrically modulated, and the IR reflection was either sent to an oscilloscope or demodulated with a lock-in amplifier.

Figure~\ref{fig:structure-setup}c shows a representative normalized IR reflection spectrum of a fabricated device without green illumination. As expected, multiple longitudinal cavity resonances are observed within the stop-band of the Bragg mirrors. One resonance marked in the main spectrum is enlarged in the inset; it is centred at $\lambda_{0}=1043.67~\mathrm{nm}$ and a Lorentzian fit gives a full width at half maximum of $\Delta\lambda=0.38~\mathrm{nm}$, corresponding to a loaded optical quality factor of $Q\approx\lambda_{0}/\Delta\lambda\approx2.8\times10^{3}$ and a Finesse of 12.7. 
Across 22 measured devices, the mean finesse was $\overline{\mathcal{F}}=12.2\pm3.6$ and the mean linewidth $\overline{\Delta\nu}=121\pm58~\mathrm{GHz}$. 
Changes in the cavity resonance wavelength are subsequently used as a sensitive probe of variations in the effective refractive index of the diamond nanobeam under green illumination.

%%%%%%%%%%%%%%%%%%%%%%%%%%%%%%%%%%%%%%%%%%%%%%%%%%%%%%%%%%%%%%%%%%%%%%%%%%%%%%%%%%%%%%%%
\subsection{Long-lived photoinduced refractive-index tuning}
\label{subsec:photorefractive_tuning}
%%%%%%%%%%%%%%%%%%%%%%%%%%%%%%%%%%%%%%%%%%%%%%%%%%%%%%%%%%%%%%%%%%%%%%%%%%%%%%%%%%%%%%%%
\begin{figure}[t]
    \centering
    \includegraphics[width=\linewidth]{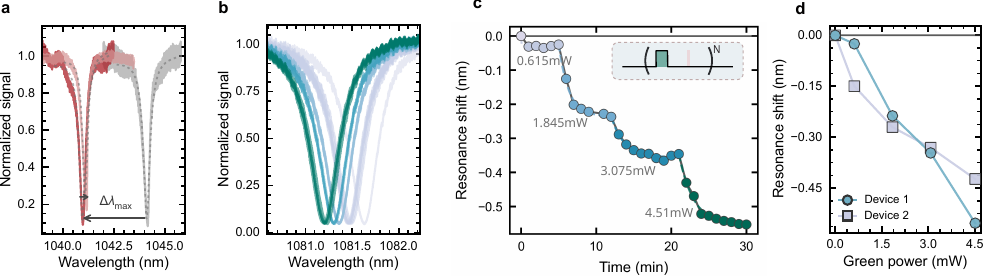}
    \caption{
    \textbf{Long-lived photorefractive tuning of diamond nanophotonic cavity resonances.}
    \textbf{a,} Normalised reflection spectra of a cavity before (grey) and after (darker red) \(5~\mathrm{h}\) of cumulative green exposure over four days. Subsequent relaxation after removal of the green light is shown in lighter red. 
    \textbf{b,} Normalised reflection spectrum of Device~1 following successive \(1~\mathrm{min}\) green-illumination intervals. Each spectrum was acquired after the green illumination had been switched off. 
    \textbf{c,} Retained resonance shift of Device~1 as a function of accumulated illumination time as the green power was increased sequentially from \(0.615\) to \(4.51~\mathrm{mW}\). The inset illustrates the measurement protocol with sequences of \(1~\mathrm{min}\) exposure at \(532~\mathrm{nm}\) followed by an IR spectrum acquisition.
    \textbf{d,} Endpoint retained resonance shifts obtained after successive green-power steps for two independent devices. 
    }
    \label{fig:photorefractive_tuning}
\end{figure}
%%%%%%%%%%%%%%%%%%%%%%%%%%%%%%%%%%%%%%%%%%%%%%%%%%%%%%%%%%%%%%%%%%%%%%%%%%%%%%%%%%%%%%%%

We first examine the slow cavity tuning that persists after the green excitation is removed. Repeated illumination of the suspended diamond cavities with \(532~\mathrm{nm}\) light produces a pronounced shift of all optical resonances towards shorter wavelengths. Figure~\ref{fig:photorefractive_tuning}a shows the largest displacement observed among the measured devices. This measurement was performed over four days, with a cumulative green-exposure time of approximately \(5~\mathrm{h}\). The initial \(1~\mathrm{h}\) exposure was performed at \(3.28~\mathrm{mW}\), while the remaining exposures were conducted predominantly at \(0.41~\mathrm{mW}\), with some variation in the applied power. 
The final resonance is detuned by \(\Delta\lambda_{\mathrm{max}}=-3.15~\mathrm{nm}\) from the initial state, corresponding to an increase in resonance frequency of approximately \(0.87~\mathrm{THz}\), which is reminiscent of recent observations by the Barclay group~\cite{Itoi2025NonvolatilePhotorefractiveDiamond,Itoi2026LargePhotorefractiveShift}. 
Using \(\Delta n_{\mathrm{eff}}\simeq n_g\Delta\lambda/\lambda_0\) \cite{Rakich2006Ultrawide} with \(n_g\approx2.39\) (obtained from mode simulation), the effective refractive index change (averaged along the cavity length) is approximately \(-7.2\times10^{-3}\).
Despite this pronounced tuning, the cavity quality factor is preserved: between the initial and maximally shifted states, the fitted linewidth and resonance contrast vary by less than \(1\%\). 
After one night without green illumination, the resonance red-shifted by approximately \(0.2~\mathrm{nm}\), i.e., only 6\% of the light-induced blue-shift. A separate device characterised in Fig.~\ref{fig:maximum_photorefractive_shift} exhibited a recovery of \(0.71~\mathrm{nm}\) over five days at room temperature. 

The temporal evolution and power dependence are investigated more thoroughly in a separate device shown in Figs.~\ref{fig:photorefractive_tuning}b and c. As illustrated schematically in the inset of Fig.~\ref{fig:photorefractive_tuning}c, the \(532~\mathrm{nm}\) elliptical focused beam was applied in successive \(1~\mathrm{min}\) intervals and switched off before each IR reflection spectrum was acquired. The IR power guided through the etched fibre was kept low at \(P_{\mathrm{IR}}=0.17~\mathrm{mW}\) to minimise IR-induced thermal nonlinearities. 
The plotted resonance shift therefore represents the quasi-permanent cavity state retained after each green exposure, free of transient effects present during illumination. Its time evolution is fitted phenomenologically using a single exponential function, and we find a monotonous increase of the time constant with increasing green power.
When similar measurements are performed on a 50~$\mu$m long cavity, the behaviour is qualitatively the same, except that the transient time scale before stabilization increases to several minutes, as shown in Fig.~\ref{fig:photorefractive_50um_comparison}. 
These observations are consistent with charge diffusion along the cavity \cite{Jayakumar2016OpticalPatterningTrappedCharge} governing the temporal evolution of refractive index change, possibly through a photorefractive effect \cite{gunter2007photorefractive} (see Discussion below). 
%The time constant at the lowest power is well below 1 min (not resolved), and they are approximately \(1.3~\mathrm{min}\) at \(1.85~\mathrm{mW}\), \(1.7~\mathrm{min}\) at \(3.07~\mathrm{mW}\), and \(2.0~\mathrm{min}\) at \(4.51~\mathrm{mW}\). 
Figure~\ref{fig:photorefractive_tuning}d compares the endpoint shifts obtained after successive power steps for two similar devices with Bragg periods differing by \(5~\mathrm{nm}\). Both devices exhibit an increasing accumulated blue shift, reaching approximately \(-0.55~\mathrm{nm}\) for Device~1 and \(-0.42~\mathrm{nm}\) for Device~2 at \(4.51~\mathrm{mW}\), with no sign of saturation. 

%%%%%%%%%%%%%%%%%%%%%%%%%%%%%%%%%%%%%%%%%%%%%%%%%%%%%%%%%%%%%%%%%%%%%%%%%%%%%%%%%%%%%%%%
\subsection{Transient response}
\label{subsec:transient_modulation}

\begin{figure*}[t]
    \centering
    \includegraphics[
        width=0.8\textwidth,
        keepaspectratio
    ]{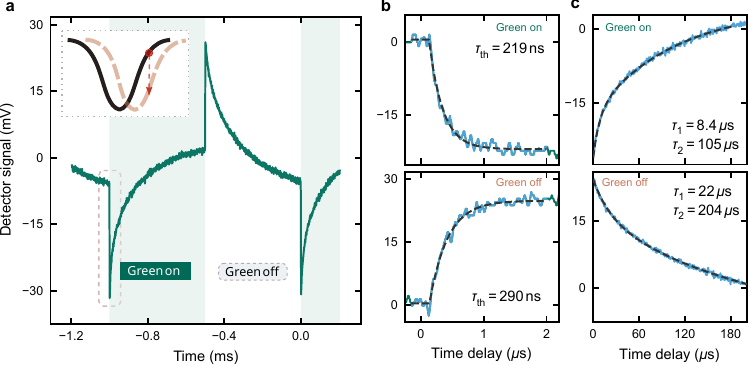}
    \caption{
    \textbf{Time-domain cavity response under square-wave green excitation.}
    \textbf{a}, Representative time trace of the reflected IR intensity for a fixed red-detuned probe laser. Upon switching on the green excitation (2.7~mW, green-shaded regions), the prompt thermo-optic redshift reduces the reflected detector signal, as illustrated in the inset. Superimposed is a blue shift on $\mu$s timescales, attributed to free-carrier plasma dispersion. 
    \textbf{b}, Zoom-in on the fast thermal and slower charge-related (\textbf{c}) responses, with mono- and bi-exponential fitting functions, respectively, plotted as dashed lines. The fitted time constants are indicated.
    }
    \label{fig:transient_response}
\end{figure*}

To resolve the fast and reversible cavity response to green light, the IR probe wavelength is fixed to the red-side slope of the resonance while the green laser intensity is square-wave modulated with a \(50\%\) duty cycle. As illustrated in the inset of Fig.~\ref{fig:transient_response}a, under this configuration, a transient cavity red-shift causes a negative change in reflectivity, and a blue-shift a positive change.
When the green illumination (2.7~mW) switches on (green shaded area in Fig.~\ref{fig:transient_response}a), a fast red-shift is observed, with a fitted time constant of a few hundred ns (Fig.~\ref{fig:transient_response}b). This time scale is fully consistent with the thermo-optic effect, as estimated using the thermal conductivity recently measured in similar diamond nanostructures \cite{goblot2026heatTransportDiamond}. The corresponding spectral shift, temperature rise, heating power, and expected absorption by the NV ensemble are estimated in Supplementary Sec.~\ref{sec:thermal_absorption_estimate}. %As expected for heat diffusion along a one-dimensional channel, for which \(\tau_{\mathrm{th}}\propto L^2\), the thermal time constant increases by a factor of approximately \(6.25\) when the suspended diamond length increases from \(20\) to \(50~\mu\mathrm{m}\) (see Fig.~\ref{fig:thermal_length_dependence}).
For a typical \(20~\mu\mathrm{m}\) cavity measured under circular green-beam excitation, the magnitude of the resonance shift corresponds to a mode-averaged temperature increase of approximately \(0.11~\mathrm{K}\) for \(1.09~\mathrm{mW}\) of incident green power at the sample plane. The corresponding heating power is approximately \(1.58~\mu\mathrm{W}\), indicating that about \(0.15\%\) of the incident power is absorbed and converted into heat. Based on the NV concentration of \(4~\mathrm{ppm}\) and the reported absorption cross-section of \(3.1\times10^{-17}~\mathrm{cm^2}\) at \(532~\mathrm{nm}\)~\cite{Wee2007}, the NV centres should absorb at most approximately \(0.054\%\) of the total incident power, some of it reemitted as photoluminescence (PL). 
Consequently, direct NV absorption alone seems insufficient to explain the observed heating, suggesting absorption by other defects, such as P1 centres (substitutional nitrogen), vacancy complexes, or surface states. %These additional absorbers may also contribute to the generation of the photo-carriers involved in the other modulation mechanisms~\cite{Barry2024SensitiveMagnetometry,Heremans2009PhotoexcitedElectrons}.

Superimposed on the fast thermal response is a slower, multi-exponential contribution of opposite sign, with fitted bi-exponential time constants ranging from a few to hundreds of microseconds (Fig.~\ref{fig:transient_response}c). The blue shift observed under illumination resembles the long-lived tuning discussed above but is reversible on these timescales: when the green excitation is switched off, the response reverses and relaxes with similar, though generally longer, time constants. The same competing fast thermo-optic and slow charge-mediated responses are observed over the full investigated green-power range, up to \(P_{\mathrm{in}}=11.89~\mathrm{mW}\) (Fig.~\ref{fig:transient_response_high_power}). 
While the fast thermal response remains on a sub-microsecond timescale, the slower dynamics become progressively faster with increasing green power (Fig.~\ref{fig:transient_timescales_power}).
This power dependence may reflect density-dependent carrier dynamics or the involvement of charge traps with different characteristic trapping and release times, as discussed below.

%%%%%%%%%%%%%%%%%%%%%%%%%%%%%%%%%%%%%%%%%%%%%%%%%%%%%%%%%%%%%%%%%%%%%%%%%%%%%%%%%%%%%%%%
%%%%%%%%%%%%%%%%%%%%%%%%%%%%%%%%%%%%%%%%%%%%%%%%%%%%%%%%%%%%%%%%%%%%%%%%%%%%%%%%%%%%%%%%

\begin{figure}[t]
    \centering
    \includegraphics[width=0.85\columnwidth,keepaspectratio]{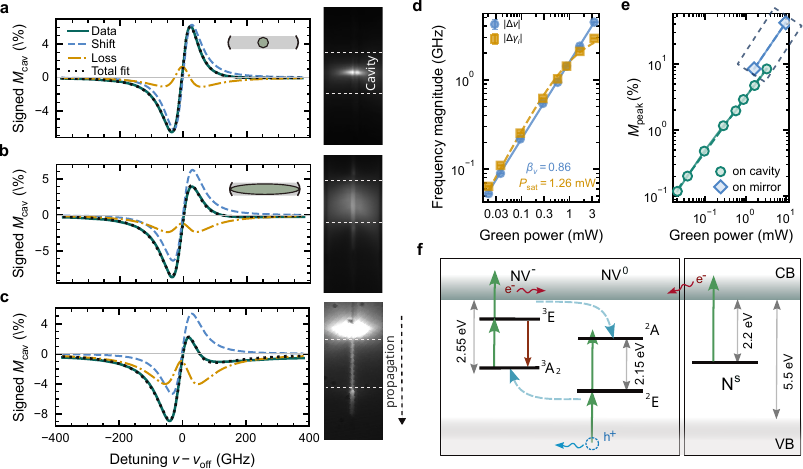}
    \caption{
    \textbf{Modulation of refractive index and absorption coefficient extracted from LIA spectra.}
    \textbf{a--c,} Contrast-normalised LIA spectra \(M_{\mathrm{cav}}\) for circular excitation of the cavity center (\(P_{\mathrm{in}}=3.53~\mathrm{mW}\)), elliptical excitation of the entire cavity (\(P_{\mathrm{in}}=3.28~\mathrm{mW}\)), and waveguide-coupled excitation (\(P_{\mathrm{in}}=1.6~\mathrm{mW}\)). The horizontal axis shows the frequency detuning \(\nu-\nu_{\mathrm{off}}\), where \(\nu_{\mathrm{off}}\) is the green-off resonance frequency for each trace. Dark-green curves show the measured response and black dotted curves the total fits. Blue dashed and orange dash-dotted curves show the fitted contributions of frequency and loss modulations, respectively. Insets indicate the illumination geometry. On the right, the acquired PL images reflect NV center emission from the suspended structure, but also from the underlying substrate.
    \textbf{d,} Magnitudes of the effective frequency modulation, \(\lvert\Delta\nu\rvert\), and internal loss modulation, \(\lvert\Delta\gamma_i\rvert\), as a function of incident green power for elliptical excitation, like in panel~\textbf{b}. For reference, the green-off total resonance linewidth is approximately \(105~\mathrm{GHz}\) (full width at half maximum). The lines show an empirical power-law fit to \(\lvert\Delta\nu\rvert\), with exponent \(\beta_{\omega}\simeq0.86\), and a saturation model fit to \(\lvert\Delta\gamma_i\rvert\).
    \textbf{e,} Peak contrast-normalised cavity modulation \(M_{\mathrm{peak}}\) as a function of incident green power. Green circles show measurements for an elliptical spot; blue diamonds for waveguide-coupled green excitation of the same cavity. The green line is a power-law fit with exponent 0.83; the blue line is a guide to the eye.
    \textbf{f,} Schematic of some possible dopant ionization processes involving \(\mathrm{NV}^{-}\), \(\mathrm{NV}^{0}\) and substitutional nitrogen \(\mathrm{N}^{\mathrm{s}}\). CB and VB: diamond conduction and valence bands. Each green arrow represents the energy of one green photon, and blue dashed arrows indicate charge-state-conversion pathways. Panel adapted from~\cite{Dhomkar2018ChargeDynamics,Itoi2025NonvolatilePhotorefractiveDiamond}.
    }
    \label{fig:modulation_depth}
\end{figure}

To obtain a more complete picture of the modulation mechanisms and assess the respective contributions of dispersive and dissipative effects, we perform wavelength-resolved measurements with a lock-in amplifier (LIA).
The green excitation is modulated at \(20~\mathrm{kHz}\), and the reflected infrared signal is demodulated by the LIA referenced to the green modulation. Each green-on or green-off interval lasts \(25~\mu\mathrm{s}\), so that the LIA response contains contributions from both mechanisms identified in Fig.~\ref{fig:transient_response}, with the slower contribution potentially attenuated and phase-delayed. 
Figures~\ref{fig:modulation_depth}a--c compare representative signed LIA spectra for different green illumination conditions depicted as insets and visible in the PL images on the right. 
The spectra are normalized to the resonance contrast and fitted using a cavity-response model that assumes a constant (green-illumination-independent) external coupling rate, as detailed in Supplementary Section~\ref{sec:supp_lia_shift_loss}. 
The main observation is the asymmetric modulation amplitude at red vs. blue detunings, not expected from a pure shift of a Lorentzian resonance; we interpret this as a signature of free-carrier-induced absorption.
The response to green light is described by a change in the resonance frequency, \(\Delta\nu\), and a simultaneous change in the internal cavity energy-decay rate, \(\Delta\gamma_i\), due to free-carrier absorption. Positive and negative values of \(\Delta\gamma_i\) correspond to broadening and narrowing, respectively. 
The fitted spectral contributions of cavity shift and loss are plotted separately in Figs.~\ref{fig:modulation_depth}a--c, and the total model reproduces the measured asymmetric spectral response. 
As the green spot is changed from a circular to an elliptical shape, the LIA spectrum becomes more asymmetric, indicating a larger relative contribution of free-carrier absorption compared to the thermal shift. %This trend is even more pronounced when green light is launched into the cavity, providing more homogeneous carrier excitation along the waveguide.  
Coupling even a small fraction of the green light into the suspended waveguide increases the interaction length and the spatial overlap between the induced free carriers and the infrared mode, providing a possible explanation for the enhanced absorption response in Fig.~\ref{fig:modulation_depth}c.

%%%%%%%%%%%%%%%%%%%%%%%%%%%%%%%%%%%%%%%%%%%%%%%%%%%%%%%%%%%%%%%%%%%%%%%%%%%%%%%%%%%%%%%%
%%%%%%%%%%%%%%%%%%%%%%%%%%%%%%%%%%%%%%%%%%%%%%%%%%%%%%%%%%%%%%%%%%%%%%%%%%%%%%%%%%%%%%%%

Fig.~\ref{fig:modulation_depth}d shows the fitted modulation amplitudes of the resonance frequency and internal loss for all green powers under an elliptical excitation spot. 
The resonance frequency modulation follows an empirical power law, \(\lvert\Delta\nu\rvert\propto P_{\mathrm{in}}^{\beta_{\omega}}\), with \(\beta_{\omega}\simeq0.86\); the absence of clear saturation is consistent with the dominant thermal origin of the index modulation. 
The internal loss modulation is better fitted with a saturation model, yielding a saturation power of 1.26~mW. 
This connects to the free-carrier generation mechanism causing absorption, as discussed below.
Note that our rate-equation estimate indicates that NV$^-$ singlet absorption (peaked at 1042~nm) is largely insufficient to explain the measured green-induced loss modulation near both \(1045~\mathrm{nm}\) and \(1081~\mathrm{nm}\) (Supplementary Section~\ref{sec:singlet_absorption_model}). 
Moreover, the metastable singlet-state population is known to relax in less than 1~$\mu$s after switching off the green light, much faster than the free-carrier dynamics seen in Fig.~\ref{fig:transient_response}c.

To illustrate the performance of the device as a photonic modulator, the effective modulation depth at the optimal IR laser detuning, \(M_{\mathrm{peak}}\), is plotted in Fig.~\ref{fig:modulation_depth}e as a function of green power for an elliptical spot (open circles). It reaches up to \(8\%\) at \(P_{\mathrm{in}}=3.28~\mathrm{mW}\) and follows an empirical power law with exponent \(\alpha=0.83\). 
For waveguide-coupled excitation of the same device and cavity resonance, \(M_{\mathrm{peak}}\) reaches approximately \(45\%\) at \(P_{\mathrm{in}}=9.6~\mathrm{mW}\), representing the largest modulation observed in this study. 
No saturation was observed for any of the measured devices over the accessible green-power range, with power-law exponents typically between \(0.8\) and \(1\).
This is consistent with the frequency modulation being dominated by the photothermal effect at 20~kHz; yet, as independently demonstrated in Fig.~\ref{fig:transient_response}, the free-carrier modulation has a similar or even larger magnitude, which would dominate in better thermalized geometries such as rib waveguides.

\section{Discussion}

The increased absorption coefficient under green illumination (Figs.~\ref{fig:modulation_depth}a-d), in conjunction with the resonance blue shift (Figs.~\ref{fig:transient_response}a,c), is the expected manifestation of the free-carrier plasma dispersion effect \cite{reed2005}, well captured by a Drude model of the linear susceptibility (see Sec.~\ref{sec:SI_transient_carrier_density}).
As illustrated in Fig.~\ref{fig:modulation_depth}f, illumination at \(532~\mathrm{nm}\) can drive charge-state conversion of NV centres and other defects and, importantly, can photoionize neutral substitutional nitrogen. Since the photon energy (\(\sim2.33~\mathrm{eV}\)) exceeds the optical ionization threshold of substitutional nitrogen (\(\sim2.2~\mathrm{eV}\)), the single-photon process \(N_{\mathrm{s}}^{0}+h\nu\rightarrow N_{\mathrm{s}}^{+}+e^{-}\) provides a possible source of mobile electrons~\cite{Manson2026SpinDependentChargeState}, scaling linearly with power. In parallel, optical cycling between NV\(^{-}\) and NV\(^{0}\) generates transient electrons or holes through ionization and recharging~\cite{Bourgeois2015PhotoelectricDetectionNV,Jayakumar2016OpticalPatterningTrappedCharge,Dhomkar2018ChargeDynamicsNV}. The photogenerated carriers can diffuse, recombine, or become trapped at NV centres, substitutional nitrogen, surface states, or other defects. Defect-to-defect charge transfer, including tunnelling between nearby nitrogen and NV centres, may additionally contribute to the subsequent charge equilibration~\cite{Manson2026SpinDependentChargeState,Alkauskas2025ChargeStateEquilibration}. 

We estimate in Sec.~\ref{sec:SI_transient_carrier_density} the free-carrier concentration required to produce the observed transient refractive-index change of approximately \(\Delta n_{\mathrm{eff}}\simeq-0.5\times10^{-6}\) for the calibrated circular-spot transient at \(P_{\mathrm{in}}=1.09~\mathrm{mW}\). Within an electron-only Drude model, this corresponds to a cavity-averaged free-electron density of approximately \(8.7\times10^{14}~\mathrm{cm^{-3}}\). This represents only about \(0.1\%\) of the nominal NV density of approximately \(7.0\times10^{17}~\mathrm{cm^{-3}}\), with even more substitutional-nitrogen defects. The required carrier population is therefore compatible with ionization or redistribution of a small fraction of the available impurities. 
The lifetime of photo-carriers created in pristine diamond by UV interband absorption was measured in Ref.~\cite{ivakin2014investigation} to be on the order of 100~ns, but it drops to below 3~ns for nitrogen impurity levels more similar to our study \cite{malinauskas2008optical} -- way faster than the response measured in Fig.~\ref{fig:transient_response}. This suggests that carriers generated by defect ionization have much longer lifetimes and deserve further investigation. 
From the measured transient resonance shift and broadening, we also estimate in Sec.~\ref{subsec:matched_mobility_estimate} a carrier mobility on the order of 10~cm$^2$V$^{-1}$s$^{-1}$ (with a large uncertainty), two or three orders of magnitude smaller than obtained from Hall-effect measurements on diamond with lower nitrogen content \cite{ishaqzai2026review}, but closer to the measured value on nitrogen-doped HPHT diamond \cite{buga2024hall}.

Another consequence of the photo-induced charge dynamics is the build-up of a quasi-permanent microscopic field distribution that may differ from the pre-illumination one. 
Based on this hypothesis, two mechanisms can explain the observed photorefractive effect. 
If we first consider our diamond as a centrosymmetric material, the DC Kerr effect predicts a quadratic dependence of $\Delta n$ on the applied electric field, with a prefactor proportional to the Kerr coefficient $n_2$, measured at sufficiently low optical frequencies \cite{almeida2017nonlinear}. 
However, $n_2$ is positive in diamond; the DC Kerr effect is expected to cause a red-shift of the resonance in our experiment. It cannot explain the permanent blue-shift. 
We therefore need to consider that NV centers, and possibly other point defects, locally break inversion symmetry in the lattice, which has been reported to give rise to a second-order nonlinear optical response \cite{Abulikemu2021SecondHarmonicDiamond,abulikemu2022temperature,Flagan2025OpticalSwitchingDiamond,sato2025ultrafast}. 
In such cases, the electro-optic (Pockels) effect can cause either a decrease or an increase in refractive index, depending on the local alignment between the applied electric field and the nonlinear response (i.e., the sign of $\chi^{(2)}$) \cite{song2026ultrafast}. 
The photorefractive effect emerges when photo-excited carriers spontaneously migrate to establish a static field, causing an illumination-dependent change in refractive index \cite{gunter2007photorefractive}.
In particular, in non-centrosymmetric crystals such as lithium niobate, photorefraction may occur even under homogeneous illumination through photovoltaic charge separation, which describes the sign-dependent preferential hopping of photogenerated carriers in a specific direction, typically along the optics axis \cite{joiner1979photorefractive}. 
Limiting our considerations to NV centers, we therefore hypothesize that a microscopic photovoltaic charge separation may favour the alignment of the local microscopic electric field from trapped carriers with the NV orientation, resulting in a net electro-optic modulation of the same sign everywhere that does not average to zero over a macroscopic volume. 
Other potential mechanisms need to break the isotropy of diamond on the scale of optical wavelengths, possibly due to the waveguide surface. 
For comparison, reproducing the largest observed long-lived blue spectral shift of \(3.15~\mathrm{nm}\) through the Pockels effect in a lithium-niobate waveguide electro-optic modulator would require approximately \(220~\mathrm{V}\) across a \(5~\mu\mathrm{m}\) electrode gap, even assuming ideal electro-optic overlap (\(\Gamma=1\)). For a more conservative overlap factor of \(\Gamma=0.5\), the estimated voltage increases to approximately \(445~\mathrm{V}\). 

\section{Conclusion}
\label{sec:conclusion}

We have demonstrated that local optical excitation can produce both long-lived refractive-index tuning and fast reversible modulation in suspended single-crystal diamond nanophotonic cavities doped with nitrogen impurities and vacancies. 
Prolonged illumination with \(532~\mathrm{nm}\) light drives the cavity resonances towards shorter wavelengths, with refractive index changes equivalent to applying several hundred volts to an ideal lithium niobate waveguide modulator, and without measurable excess loss. 
This quasi-permanent shift is consistent with a photorefractive effect, possibly driven by a microscopic photovoltaic charge separation or by surface charges, in conjunction with the second-order nonlinearity induced by NV centers and other non-centrosymmetric defects. 
Time-resolved measurements further reveal a transient and reversible response that is the sum of a photo-thermal red-shift and a free-carrier plasma dispersion effect. 
The latter manifests as a simultaneous blue-shift and broadening of the resonance, in line with expectations from the Drude model. 
When coupling a small fraction of the green light into the diamond waveguide, the modulation depth reaches close to 50\%, establishing our optical method as a viable route to active diamond photonics.

The present measurements establish that doped diamond nanostructures can support rich and strong optical responses that lay a basis for fast reversible modulation and long-lived post-fabrication tuning. 
Further improvements in green-light coupling and cavity design will increase the spatial overlap with the cavity mode, enabling larger modulation depths at lower incident powers. 
Improved control of the defect and surface environment may also enhance the stability and reproducibility of the photo-induced charge response~\cite{Kumar2024NVSurfacePassivation,Hauf2011ChemicalControlNV}. 
The integration of metal electrodes around the waveguide may enable GHz-rate electro-optic modulation. 
Together, these developments contribute to developing programmable and reconfigurable quantum circuits \cite{raniwala2026full,aharonovich2026}, with applications such as on-chip single-photon waveform shaping \cite{wu2025a}, agile and dynamically addressable quantum sensors \cite{weng2026} or quantum-memory arrays \cite{zhang2024}, and microwave-to-optical quantum transducers and detectors \cite{li2017quantum,liu2021one,woodman2023}. 

\subsection{Funding}
This project has received funding from the Swiss National Science Foundation through grants 198898 and 204036. V. G. acknowledges support from the Swiss National Science Foundation through the Ambizione Fellowship, grant 216406. 

\subsection{Data Availability}
All data underlying the claims in this manuscript are available from the authors upon reasonable request. 

\subsection{Author Contribution}
Y. Z., with assistance from A. A., designed and fabricated the samples.  Y. Z., with assistance from V. G. and C. G., performed the measurements and analyzed the data. T. L.T., N. Q., V. G. and C. G. supervised the project. Y. Z., V. G. and C. G. wrote the manuscript, and all authors reviewed it and provided their feedback.

\subsection{Competing Interest}
The authors declare no competing interests. 

\bibliographystyle{unsrt}
\bibliography{references}

%%%%%%%%%%%%%%%%%%%%%%%%%%%%%%%%%%%%%%%%%%%%%%%%%%%%%%%%%%%%%%%%%%%%%%%%%%%%%%%%
%%%%%%%%%%%%%%%%%%%%%%%%%%%%%%%%%%%%%%%%%%%%%%%%%%%%%%%%%%%%%%%%%%%%%%%%%%%%%%%%
\newpage

\setcounter{section}{0}
\setcounter{subsection}{0}

\renewcommand{\thesection}{S\arabic{section}}
\renewcommand{\thesubsection}{\thesection.\arabic{subsection}}
\setcounter{secnumdepth}{2}

\makeatletter
\renewcommand{\p@subsection}{}
\renewcommand{\p@subsubsection}{}
\makeatother

\renewcommand{\figurename}{\textbf{Supplementary Figure}}
\renewcommand{\tablename}{\textbf{Supplementary Table}}
\renewcommand{\thefigure}{S\arabic{figure}}
\renewcommand{\thetable}{S\arabic{table}}

\setcounter{figure}{0}
\setcounter{table}{0}

\begin{center}
    {\Large SUPPLEMENTARY MATERIAL}\\~\\
    \textbf{Optically-Induced Modulation and Programming of Diamond Photonic Cavities}\\~\\
    Yuchun Zhu, Amirali Arabmoheghi, Nicolas Le Thomas, Niels Quack, Valentin Goblot, and Christophe Galland
\end{center}

%%%%%%%%%%%%%%%%%%%%%%%%%%%%%%%%%%%%%%%%%%%%%%%%%%%%%%%%%%%%%%%%%%%%%%%%%%%%%%%%
\section{Design parameters of the asymmetric Fabry--Pérot nanobeam cavity}
\label{sec:fp_design_parameters}

The nominal design parameters of the Fabry--Pérot nanobeam cavities are summarized in Table~\ref{tab:fp_design_parameters}.  The Bragg period is swept across the chip area to achieve the targeted stop band in the range from 1000 to 1100~nm.

\begin{table*}[h]
    \centering
    \caption{Nominal design parameters of the asymmetric Fabry--Pérot nanobeam cavity. The mirror lengths refer to the uniform Bragg sections and exclude the tapered transitions. }
    \label{tab:fp_design_parameters}

    \renewcommand{\arraystretch}{1.15}
    \setlength{\tabcolsep}{10pt}

    \begin{tabular}{@{}l c l@{}}
        \toprule
        Parameter & Symbol & Nominal value \\
        \midrule
        \multicolumn{3}{@{}l}{\textit{Waveguide cross-section}} \\
        \addlinespace[2pt]

        Target wavelength
        & \(\lambda_0\)
        & \(1042~\mathrm{nm}\) \\

        Diamond thickness
        & \(t\)
        & \(340\pm20~\mathrm{nm}\) \\

        Straight-waveguide width
        & \(w\)
        & \(437~\mathrm{nm}\) \\

        \addlinespace[5pt]
        \multicolumn{3}{@{}l}{\textit{Bragg-mirror geometry}} \\
        \addlinespace[2pt]

        Bragg period
        & \(\Lambda\)
        & \(304~\mathrm{nm}\) \\

        Sidewall-corrugation amplitude
        & \(k\)
        & \(100~\mathrm{nm}\) (each side) \\

        Minimum and maximum widths
        & \(w_{\min},\,w_{\max}\)
        & \(237,\ 637~\mathrm{nm}\) \\

        Input-mirror periods
        & \(N_{\mathrm{B},1}\)
        & \(30\) \\

        End-mirror periods
        & \(N_{\mathrm{B},2}\)
        & \(60\) \\

        Transition periods
        & \(N_{\mathrm{t}}\)
        & \(5\) per transition \\

        \addlinespace[5pt]
        \multicolumn{3}{@{}l}{\textit{Cavity and coupling geometry}} \\
        \addlinespace[2pt]

        Nominal cavity lengths
        & \(L_{\mathrm{cav}}\)
        & \(20,\ 50,\ 60~\mu\mathrm{m}\) \\

        Target input-mirror reflectivity
        & \(R_1\)
        & \(81\%\) \\

        Target end-mirror reflectivity
        & \(R_2\)
        & \(99\%\) \\

        Inverse-taper length
        & \(L_{\mathrm{taper}}\)
        & \(20~\mu\mathrm{m}\) \\

        \bottomrule
    \end{tabular}
\end{table*}

%%%%%%%%%%%%%%%%%%%%%%%%%%%%%%%%%%%%%%%%%%%%%%%%%%%%%%%%%%%%%%%%%%%%%%%%%%%%%%%%
%\subsection{Relaxation of the quasi-permanent cavity shift}
\section{Supplementary Figures}

\begin{figure}[H]
    \centering
    \includegraphics[width=0.8\textwidth]{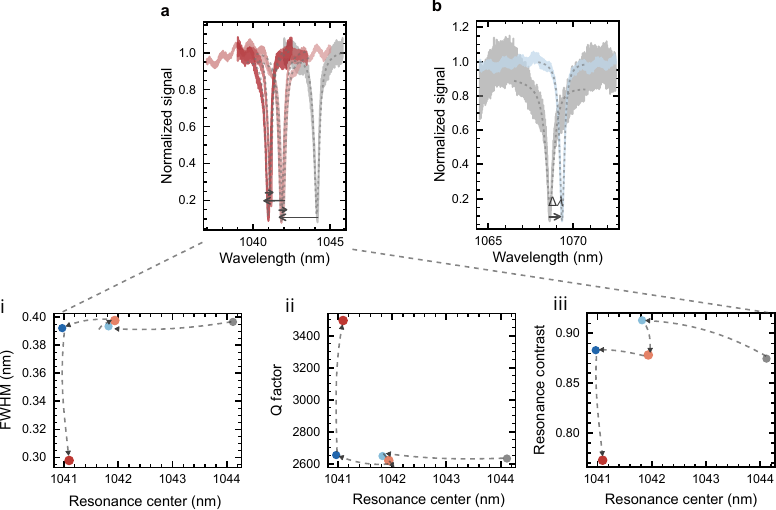}
    \caption{
    \textbf{Large photoinduced cavity blue shift and subsequent relaxation.}
    \textbf{a,} Normalised infrared reflection spectra of Device~5 at selected stages of green-induced tuning and subsequent recovery. The maximum blue shift relative to the initial resonance reaches \(\Delta\lambda_{\mathrm{max}}=-3.15~\mathrm{nm}\) after approximately \(5~\mathrm{h}\) of cumulative green illumination distributed over four days. Dotted curves show fitted resonance profiles.
    \textbf{i--iii,} Fitted full width at half maximum (FWHM), quality factor \(Q\), and resonance contrast, respectively, plotted against the fitted resonance centre for the spectra in \textbf{a}. Marker colours identify the corresponding spectra. The initial and maximally blue-shifted states have comparable linewidths, quality factors, and contrasts, indicating that the large spectral shift is not accompanied by a substantial deterioration of these resonance properties. The highest quality factor occurs in the final spectrum of the measurement sequence.
    \textbf{b,} Partial recovery of the photoinduced shift in another cavity. The spectrum recorded in the shifted state (light grey) is compared with that measured five days later (light blue). The resonance shifts back towards its initial wavelength by approximately \(0.71~\mathrm{nm}\), as indicated by the arrow, showing that the photoinduced shift relaxes over several days. The larger noise in the grey trace results from suboptimal fibre-to-waveguide coupling.
    }
    \label{fig:maximum_photorefractive_shift}
\end{figure}

%%%%%%%%%%%%%%%%%%%%%%%%%%%%%%%%%%%%%%%%%%%%%%%%%%%%%%%%%%%%%%%%%%%%%%%%%%%%%%%%
%\subsection{Comparison of remanent tuning in cavities of different lengths} \label{subsec:cavity_length_photorefractive_dynamics}

\begin{figure}[H]
    \centering
    \includegraphics[width=0.6\linewidth]{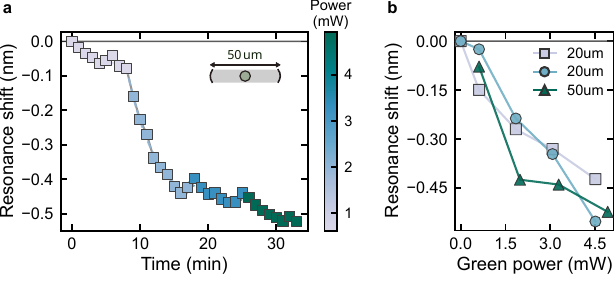}
    \caption{
    Green-induced remanent tuning in cavities of different lengths.
    \textbf{a,} Retained spectral shift of a \(50~\mu\mathrm{m}\)-long cavity following successive green-exposure intervals. The colour scale indicates the applied green power. Exponential fits yield time constants of \(3.28\), \(4.35\), \(5.37\), and \(3.73~\mathrm{min}\) at green powers of \(0.61\), \(1.98\), \(3.28\), and \(4.92~\mathrm{mW}\), respectively, almost four times longer than the values extracted from the shorter \(20~\mu\mathrm{m}\) cavities, which is consistent with a diffusive process along the waveguide.
    \textbf{b,} Endpoint shifts after successive power steps for two \(20~\mu\mathrm{m}\)-long cavities (Devices~1 and 2) and one \(50~\mu\mathrm{m}\)-long cavity (Device~3), showing no saturation.
    }
    \label{fig:photorefractive_50um_comparison}
\end{figure}

%%%%%%%%%%%%%%%%%%%%%%%%%%%%%%%%%%%%%%%%%%%%%%%%%%%%%%%%%%%%%%%%%%%%%%%%%%%%%%%%%%%%%%%%%%%%%%%%%%%%%%%%%%%%%%%
%%%%%%%%%%%%%%%%%%%%%%%%%%%%%%%%%%%%%%%%%%%%%%%%%%%%%%%%%%%%%%%%%%%%%%%%%%%%%%%%
\section{Green-power dependence of the transient response and characteristic timescales}
\label{sec:transient_power_dependence}
\label{subsec:supp_transient_timescales}

\begin{figure}[H]
\centering
\includegraphics[
width=0.9\textwidth,
keepaspectratio
]{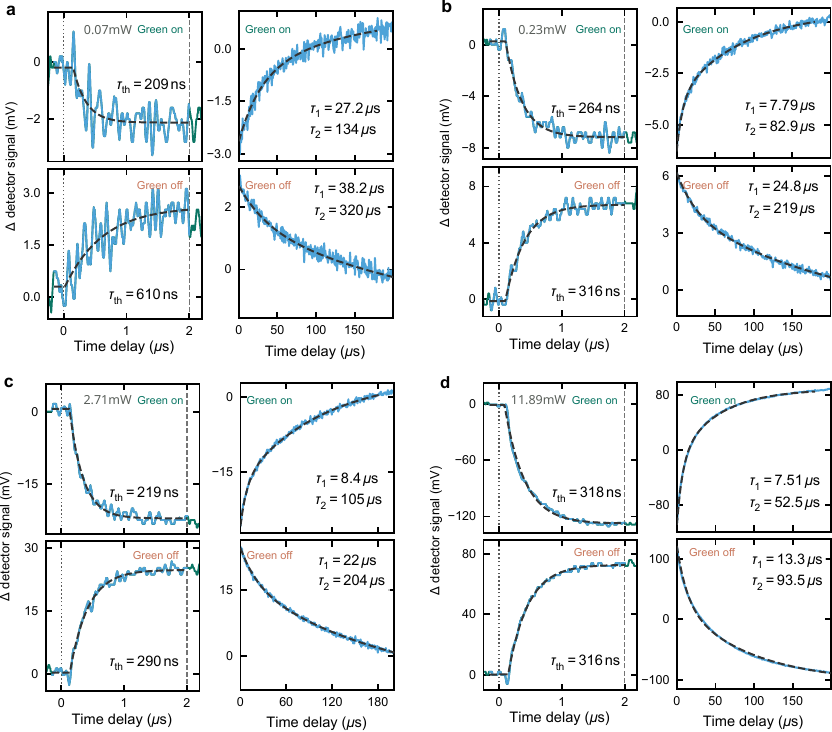}
\caption{
Green-power dependence of the transient cavity response. Representative green-on and green-off switching edges measured at sample-plane powers of \textbf{a,} \(0.07~\mathrm{mW}\), \textbf{b,} \(0.23~\mathrm{mW}\), \textbf{c,} \(2.71~\mathrm{mW}\), and \textbf{d,} \(11.89~\mathrm{mW}\). The left-hand plots resolve the fast thermo-optic response, while the right-hand plots show the subsequent slow evolution. Blue curves show the measured detector signal and black dashed curves show the fits.
}
\label{fig:transient_response_high_power}
\end{figure}

The transient cavity response was measured under square-wave green excitation at sample-plane powers \(P_{\mathrm{in}}\) ranging from \(0.0738~\mathrm{mW}\) to \(11.89~\mathrm{mW}\). As shown by the representative traces in Fig.~\ref{fig:transient_response_high_power}, each green-on and green-off switching edge contains a fast sub-microsecond response followed by a slower evolution in the opposite direction. This qualitative behaviour persists over the complete investigated power range.

\begin{figure}[H]
\centering
\includegraphics[
width=0.7\columnwidth,
keepaspectratio
]{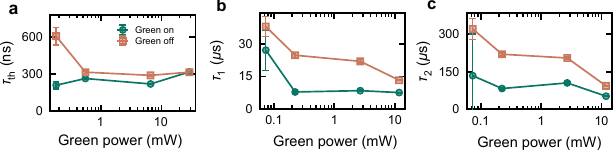}
\caption{
Green-power dependence of the characteristic cavity-response timescales extracted from the transient measurements. \textbf{a,} Fast thermo-optic time constant, \(\tau_{\mathrm{th}}\), obtained from single-exponential fits to the initial response following the green-on and green-off switching edges. \textbf{b,c,} Slow time constants, \(\tau_1\) and \(\tau_2\), respectively, obtained from bi-exponential fits to the subsequent evolution. Green circles and orange squares denote values extracted following the green-on and green-off edges, respectively. Error bars indicate the uncertainties obtained from the fits, and lines connecting the data points are guides to the eye.
}
\label{fig:transient_timescales_power}
\end{figure}

To quantify the dynamics, the initial fast response was fitted with a single-exponential function to obtain the thermo-optic time constant \(\tau_{\mathrm{th}}\), while the subsequent slower evolution was described phenomenologically by a bi-exponential function with time constants \(\tau_1\) and \(\tau_2\). Their power dependence is summarised in Fig.~\ref{fig:transient_timescales_power}.

The fitted thermo-optic time constant \(\tau_{\mathrm{th}}\) depends only weakly on \(P_{\mathrm{in}}\) and remains in the sub-microsecond regime throughout the investigated power range. In contrast, \(\tau_1\) and \(\tau_2\) occur on timescales of tens to hundreds of microseconds and generally decrease as \(P_{\mathrm{in}}\) is increased. The slow relaxation is systematically longer following the green-off edge than following the green-on edge, particularly at low power, indicating that the build-up and relaxation of the photo-induced response do not follow identical dynamics. Together with the increasing response amplitude, these measurements show that increasing the green power primarily strengthens and accelerates the slow charge-mediated contribution without changing its clear temporal separation from the fast thermo-optic response.

\section{Modeling the wavelength-resolved LIA response}
\label{sec:supp_lia_shift_loss}

To separate the green-induced cavity-frequency shift from the change in internal cavity loss, we performed wavelength-resolved lock-in amplifier (LIA) measurements under \(532~\mathrm{nm}\) excitation modulated at \(20~\mathrm{kHz}\).
%The measured LIA quadratures were corrected for the detector gain and LIA sensitivity, followed by application of a common phase rotation. 
We apply a phase rotation to the measured LIA quadratures \(X, Y\), to align the phase of the signal and modulation reference. The resulting signed quadrature is denoted by \(X_{\mathrm{rot}}\).

To compare measurements with different levels of reflection signal, \(X_{\mathrm{rot}}\) is normalized to the reflection contrast of the corresponding green-off cavity resonance. Noting the steady-state measured reflection
\begin{equation}
V(\lambda)
= V_\infty \mathcal{R}(\lambda)
\end{equation}
with \(V_\infty\) the maximum, off-resonant signal and \(\mathcal{R}(\lambda)\) the normalized cavity reflectivity, the LIA response \(X_{\mathrm{rot}}\) can be expressed as
\begin{equation}
X_{\mathrm{rot}}(\lambda)
=
A
\left[
V_{\mathrm{on}}(\lambda)
-
V_{\mathrm{off}}(\lambda)
\right].
\label{eq:lia_response}
\end{equation}
In the expression above, we have introduced the effective green-on and green-off reflection signal \(V_{\mathrm{on}}, V_{\mathrm{off}}\), \(\eta\) is the LIA sensitivity and \(A\) accounts for the fact that the LIA response corresponds to the first harmonic of the time-dependent signal. For a modulated signal corresponding to a perfect square wave, with a \(50\%\) duty cycle,
\begin{equation}
A
=
\frac{\sqrt{2}}{\pi}
\simeq
0.450.
\label{eq:supp_lia_scale_factor}
\end{equation}
In practice, the signal is not exactly square, and \(A\) is calibrated experimentally.

We also introduce the reflectivity contrast \(C = 1 - \mathcal{R}(\lambda_0)\). From Eq.~\eqref{eq:lia_response}, the maximum absolute modulation amplitude is bounded by \( \left| X_{\mathrm{rot}}(\lambda) \right| \leq \eta A V_\infty C\). We thus define the contrast-normalized cavity modulation as 
\begin{equation}
M_{\mathrm{cav}}(\lambda)
=
\frac{X_{\mathrm{rot}}(\lambda)}
{\eta A V_\infty C} = \frac{V_{\mathrm{on}}(\lambda) - V_{\mathrm{off}}(\lambda)}{V_\infty C},
\label{eq:contrast_normalised_cavity_modulation}
\end{equation}
This is the quantity plotted in Figs.~\ref{fig:modulation_depth}a-c. 

To model the LIA spectrum, we first model the steady-state reflection spectrum. Because the infrared light enters and leaves the cavity through the same waveguide port, the normalized reflected optical power is described by the one-port cavity-reflection model:
\begin{equation}
\mathcal{R}(\nu; \nu_0, \gamma_i, \gamma_e) = 1 - \frac{\gamma_i \gamma_e}{(\nu - \nu_0)^2 + [ (\gamma_i + \gamma_e) / 2]^2}
\label{eq:reflectivity_dip}
\end{equation}
where \(\nu_0 = h / \lambda_0\) is the cavity resonance frequency and \(\gamma_i, \gamma_e\) are the linewidth contribution from, respectively, the internal losses and coupling to the external waveguide.

The normalized, effective green-off and green-on cavity responses are written as
\begin{equation}
\begin{aligned}
V_{\mathrm{off}}(\nu)
&=
\mathcal{R}\left(
\nu;
\nu_{0},
\gamma_i,
\gamma_e
\right),
\\
V_{\mathrm{on}}(\nu)
&=
\mathcal{R}\left(
\nu;
\nu_{0}+\Delta\nu,
\gamma_i+\Delta\gamma_i,
\gamma_e
\right).
\end{aligned}
\label{eq:supp_green_off_on_responses}
\end{equation}
Here, \(\Delta\nu\) is the green-induced resonance-frequency shift and \(\Delta\gamma_i\) is the corresponding change in the internal decay rate (attributed to free-carrier absorption). The external loss rate \(\gamma_e\) is considered unaffected by the presence of greeen excitation. %A small bounded adjustment of the green-off resonance frequency, \(\nu_{0,\mathrm{eff}}\), is allowed for each measured trace, due to possible photorefractive shifts.
The normalized phase-rotated LIA signal is then fitted as:
\begin{equation}
X_{\mathrm{rot}}^{\mathrm{fit}}(\nu)
=
V_{\mathrm{on}}(\nu)
-
V_{\mathrm{off}}(\nu)
.
\label{eq:supp_finite_difference_lia_model}
\end{equation}
%Small residual offsets and slopes are included as nuisance parameters in the numerical fitting but are omitted from Eq.~\eqref{eq:supp_finite_difference_lia_model} for clarity. 
The five parameters \(\nu_{0}, \gamma_i, \gamma_e, \Delta\nu\) and \(\Delta\gamma_i\) are extracted simultaneously by nonlinear least-squares fitting.

Finally, to show the two fitted contributions from \(\Delta\nu\) and \(\Delta\gamma_i\) separately, 
we note that for small \(\Delta \nu, \Delta \gamma\), these two contributions can be considered as independent:
\begin{equation}
X_{\mathrm{rot}}^{\mathrm{fit}}(\nu)
\simeq \Delta \nu \frac{\partial \mathcal{R}}{\partial \nu_0} + \Delta \gamma_i \frac{\partial \mathcal{R}}{\partial \gamma_i}
\end{equation}
In practice, in Figs.~\ref{fig:modulation_depth}a-c, the two contributions are computed as
\begin{align}
    X_{\mathrm{rot}}^{\mathrm{shift}} &= \mathcal{R}\left(\nu; \nu_{0}+\Delta\nu, \gamma_i, \gamma_e \right) - \mathcal{R}\left(\nu; \nu_{0}, \gamma_i, \gamma_e \right) \\
    X_{\mathrm{rot}}^{\mathrm{loss}} &= \mathcal{R}\left(\nu; \nu_{0}, \gamma_i + \Delta \gamma_i, \gamma_e \right) - \mathcal{R}\left(\nu; \nu_{0}, \gamma_i, \gamma_e \right) 
\end{align}
Their sum gives the total fitted response. The frequency shift primarily produces the dispersive contribution, whereas the internal-loss change modifies the width and depth of the cavity response, and is responsible for the asymmetry of the LIA trace.

\section{Characterization of the green excitation intensity profile}
\label{sec:green_excitation_profile}

The spatial profile of the focused \(532~\mathrm{nm}\) excitation beam was characterized from camera images recorded at the sample plane. A background image acquired without green illumination was subtracted, and the resulting intensity distribution was fitted with a rotated two-dimensional Gaussian,
\begin{equation}
    S(x,y)
    =
    S_0
    \exp\left[
        -2\left(
            \frac{x'^2}{w_{\perp}^2}
            +
            \frac{y'^2}{w_{\parallel}^2}
        \right)
    \right]
    +S_{\mathrm{bg}},
    \label{eq:si_rotated_gaussian}
\end{equation}
where \(w_{\perp}\) and \(w_{\parallel}\) are the fitted \(1/e^2\) intensity radii across and along the nanobeam, respectively, and the rotated coordinates \(x'\) and \(y'\) account for the orientation of the excitation spot.

\begin{figure}[H]
    \centering
    \includegraphics[width=0.8\linewidth]{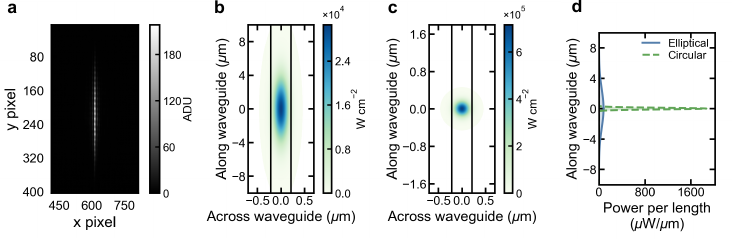}
    \caption{
    Characterisation of the green excitation profile and its geometrical overlap with the suspended nanobeam.
    (a) Background-subtracted camera image of the elliptical green illumination spot.
    (b) Reconstructed elliptical power-density distribution obtained from a rotated two-dimensional Gaussian fit. The black lines indicate the nominal boundaries of the \(437~\mathrm{nm}\)-wide nanobeam.
    (c) Modelled circular reference profile constructed using \(w_{\mathrm{circ}}=w_{\perp}\). Panels (b) and (c) use different spatial ranges and colour scales.
    (d) Power geometrically intercepted per unit nanobeam length. The circular excitation is more localised, whereas the elliptical excitation illuminates a larger fraction of the \(20~\mu\mathrm{m}\)-long cavity region.
    }
    \label{fig:green_excitation_profile}
\end{figure}

The absolute green power was measured before the microscope objective and corrected using the independently measured objective transmission, \(T_{\mathrm{obj}}=0.41\). The power incident at the sample plane was therefore \(P_{\mathrm{in}}=0.41P_{\mathrm{before\,obj}}\). The fitted spatial profile was normalised to this power to reconstruct the incident power-density distribution,
\begin{equation}
    I_{\mathrm{ell}}(u,v)
    =
    \frac{2P_{\mathrm{in}}}
         {\pi w_{\perp}w_{\parallel}}
    \exp\left[
        -2\left(
            \frac{v^2}{w_{\perp}^2}
            +
            \frac{u^2}{w_{\parallel}^2}
        \right)
    \right],
    \label{eq:si_elliptical_power_density}
\end{equation}
where \(u\) and \(v\) denote the coordinates along and across the nanobeam. The reconstructed elliptical profile is shown in Fig.~\ref{fig:green_excitation_profile}b.

For comparison, a circular reference profile was constructed using \(w_{\mathrm{circ}}=w_{\perp}\),
\begin{equation}
    I_{\mathrm{circ}}(u,v)
    =
    \frac{2P_{\mathrm{in}}}
         {\pi w_{\mathrm{circ}}^2}
    \exp\left[
        -2\frac{u^2+v^2}{w_{\mathrm{circ}}^2}
    \right].
    \label{eq:si_circular_power_density}
\end{equation}

This reference represents the expected excitation profile without expansion along the nanobeam and therefore provides a comparison between localised circular illumination and spatially extended elliptical illumination.
The fitted \(1/e^2\) intensity radii were \(w_{\perp}=0.38~\mu\mathrm{m}\) across the nanobeam and \(w_{\parallel}=8.92~\mu\mathrm{m}\) along the nanobeam. Because a fraction of the incident beam extends beyond the suspended structure, the power geometrically intercepted by the nanobeam is smaller than the total
sample-plane power. It was estimated by integrating the reconstructed power density over a rectangular region representing the \(w=0.437~\mu\mathrm{m}\)-wide and \(L_{\mathrm{wg}}=20~\mu\mathrm{m}\)-long cavity section:
\begin{equation}
    P_{\mathrm{wg}}
    =
    \int_{-L_{\mathrm{wg}}/2}^{L_{\mathrm{wg}}/2}
    \int_{-w/2}^{w/2}
    I(u,v)\,\mathrm{d}v\,\mathrm{d}u.
    \label{eq:si_intercepted_power}
\end{equation}
The corresponding geometrical overlap fraction is defined as
\begin{equation}
    f_{\mathrm{ov}}
    =
    \frac{P_{\mathrm{wg}}}{P_{\mathrm{in}}}.
    \label{eq:si_geometrical_overlap}
\end{equation}
For the elliptical Gaussian profile, this evaluates analytically to
\begin{equation}
    f_{\mathrm{ov,ell}}
    =
    \operatorname{erf}\left(
        \frac{L_{\mathrm{wg}}}{\sqrt{2}w_{\parallel}}
    \right)
    \operatorname{erf}\left(
        \frac{w}{\sqrt{2}w_{\perp}}
    \right)
    \simeq 0.730.
    \label{eq:si_elliptical_overlap}
\end{equation}
The corresponding circular reference profile gives
\begin{equation}
    f_{\mathrm{ov,circ}}
    =
    \operatorname{erf}\left(
        \frac{L_{\mathrm{wg}}}{\sqrt{2}w_{\mathrm{circ}}}
    \right)
    \operatorname{erf}\left(
        \frac{w}{\sqrt{2}w_{\mathrm{circ}}}
    \right)
    \simeq 0.749.
    \label{eq:si_circular_overlap}
\end{equation}
Thus, approximately \(73\%\) and \(75\%\) of the sample-plane power are
geometrically incident on the defined cavity region for the elliptical
and circular profiles, respectively. For example, at
\(P_{\mathrm{in}}=3~\mathrm{mW}\), these fractions correspond to
\(P_{\mathrm{wg,ell}}\simeq2.19~\mathrm{mW}\) and
\(P_{\mathrm{wg,circ}}\simeq2.25~\mathrm{mW}\). The similar total overlap
fractions do not imply similar spatial excitation: the circular profile
concentrates the power locally, whereas the elliptical profile distributes
it along a substantially larger fraction of the cavity.

The longitudinal distribution of the intercepted power was calculated as
\begin{equation}
    p(u)
    =
    \int_{-w/2}^{w/2}
    I(u,v)\,\mathrm{d}v.
    \label{eq:si_power_per_length}
\end{equation}
As shown in Fig.~\ref{fig:green_excitation_profile}d, the circular profile
produces a higher local peak, whereas the elliptical profile distributes
the excitation over a larger fraction of the cavity. The calculated
\(P_{\mathrm{wg}}\) represents only the geometrical overlap between the
incident beam and the projected nanobeam area; it does not account for
reflection, scattering, or absorption within the diamond.

\section{Refractive-index change and Drude-equivalent carrier estimate}
\label{sec:transient_physical_estimates}

\begin{figure}[!htbp]
    \centering
    \includegraphics[width=0.98\linewidth]{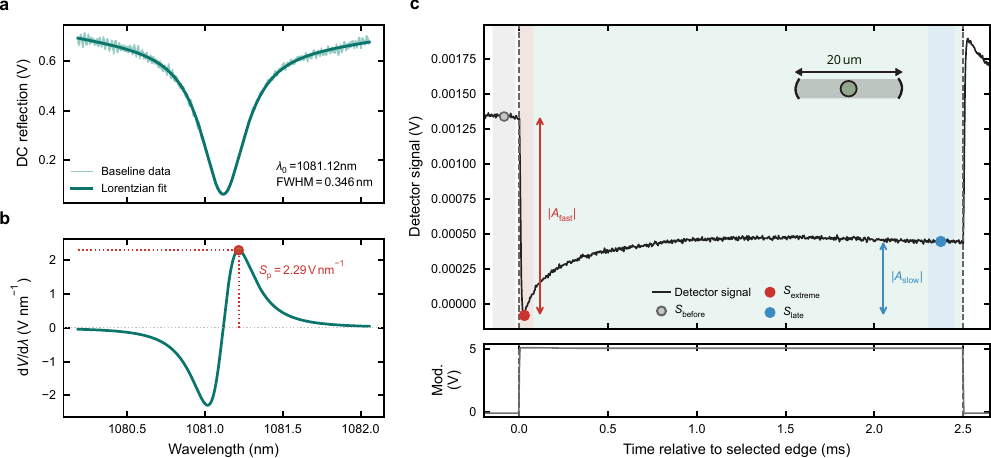}
    \caption{
    Calibration procedure used to convert the measured transient detector signal into a cavity-resonance shift.
    \textbf{a,} Green-off reflection spectrum of the \(20~\mu\mathrm{m}\)-long cavity and corresponding Lorentzian fit. The fitted resonance is centred at \(\lambda_{0}=1081.12~\mathrm{nm}\) with a linewidth of \(0.346~\mathrm{nm}\).
    \textbf{b,} Wavelength derivative of the fitted Lorentzian resonance. The red marker indicates the maximum positive slope on the right-hand side of the resonance, \(S_{\mathrm{p}}=2.29~\mathrm{V\,nm^{-1}}\), used to convert the transient detector-voltage change into an equivalent wavelength shift.
    \textbf{c,} Representative transient measured with the IR probe fixed at \(\lambda_{\mathrm{p}}=1081.26~\mathrm{nm}\) under circular \(532~\mathrm{nm}\) excitation modulated at \(200~\mathrm{Hz}\), with a calibrated sample-plane green power of \(P_{\mathrm{in}}=1.09~\mathrm{mW}\). The grey, red and blue markers denote \(S_{\mathrm{before}}\), \(S_{\mathrm{extreme}}\) and \(S_{\mathrm{late}}\), respectively. The fast amplitude is defined as
    \(\lvert A_{\mathrm{fast}}\rvert
    =\lvert S_{\mathrm{extreme}}-S_{\mathrm{before}}\rvert\),
    while the subsequent slow relaxation is quantified by
    \(\lvert A_{\mathrm{slow}}\rvert
    =\lvert S_{\mathrm{extreme}}-S_{\mathrm{late}}\rvert\).
    The lower panel shows the modulation reference used to identify the switching edges.
    }
    \label{fig:slow_fast_extraction_diagnostic}
\end{figure}

In this section, we compute order of magnitude estimates for the optical absoprtion and free-carrier plasma dispersion corresponding, respectively, to the observed fast and slow variations in the transient response. 

Fig.~\ref{fig:slow_fast_extraction_diagnostic}(c) shows the transient response, when turning green excitation on, for a \(20~\mu\mathrm{m}\) cavity under circular \(532~\mathrm{nm}\) excitation modulated at \(200~\mathrm{Hz}\), with a calibrated sample-plane green power of \(P_{\mathrm{in}}=1.09~\mathrm{mW}\). The IR probe is red-detuned from the resonance. As discussed in the main text, as the green laser is switched on, the transient response shows a fast thermal red shift followed by a slower charge-mediated blue shift. 

%Note that the circular beam is preferred for these estimates because its localized absorption-induced heating is easier to model. For the elliptical beam geometry used in Fig.~\ref{fig:transient_response}, the fast and slow components of the transient response are roughly equal to each other in magnitude, so we can safely assume that the order-of-magnitude estimate of free carrier density derived below remain valid for elliptical beam excitation.

% Prvious version:
%The clean transient response shown in Fig.~\ref{fig:slow_fast_extraction_diagnostic} contains a fast thermal red shift followed by a slower charge-mediated blue shift. For the quantitative thermal calibration, we use a \(20~\mu\mathrm{m}\) cavity measured under circular \(532~\mathrm{nm}\) excitation modulated at \(200~\mathrm{Hz}\), with a calibrated sample-plane green power of \(P_{\mathrm{in}}=1.09~\mathrm{mW}\). The circular beam is preferable for this estimate because its localised heating profile is easier to model. The slow blue-shift amplitude in the clean transient shown in the main text is of the same order of magnitude as the fast thermal red-shift amplitude. We therefore use the reliably calibrated thermal spectral-shift magnitude as an order-of-magnitude estimate of the carrier-induced blue shift, rather than performing a second absolute calibration with the elliptical-beam transient.

\subsection{Spectral-shift and effective-index calibration}
\label{subsec:transient_index_change}

We first convert the the measured change in reflection signal into its corresponding wavelength shift, and in turn, change of refractive index.
The steady-state reflection spectrum (without green excitation), Fig.~\ref{fig:slow_fast_extraction_diagnostic}(a), was fitted locally with a Lorentzian dip to determine the resonance depth, wavelength and linewidth (see section~\ref{subsec:matched_mobility_estimate} below for details).
In the transient experiments, the IR probe is taken to lie at the maximum slope of the resonance dip. For the cavity measured in Fig.~\ref{fig:slow_fast_extraction_diagnostic}, the value of the maximum slope has been extracted form the derivative of the fitted resonance (Fig.~\ref{fig:slow_fast_extraction_diagnostic}(b)):
\begin{equation}
\left|S_{\mathrm{p}}\right|
\simeq
2.3~\mathrm{V\,nm^{-1}}.
\label{eq:circular_probe_slope}
\end{equation}
This slope gives a direct conversion from a change of reflection signal into resonance shift. 
In Fig.~\ref{fig:slow_fast_extraction_diagnostic}, the measured amplitude of the fast component of the transient response is \(\lvert A_{\mathrm{fast}}\rvert=1.42~\mathrm{mV}\), corresponding to a line shift
\begin{equation}
\left|\Delta\lambda_{\mathrm{th}}\right|
=
\frac{\lvert A_{\mathrm{fast}}\rvert}{\lvert S_{\mathrm{p}}\rvert}
\simeq
0.62~\mathrm{pm}.
\label{eq:transient_wavelength_shift}
\end{equation}
The sign of the detector response and the probe position identify this component as a red shift as green excitation is switched on, \(\Delta\lambda_{\mathrm{th}}\simeq-0.62~\mathrm{pm}\) (and a blue shift of equal magnitude when green is switched off).

For a perturbation that does not change the physical cavity length, the corresponding absolute effective-index change, averaged over the cavity length, is~\cite{Rakich2006Ultrawide}
\begin{equation}
\Delta n_{\mathrm{eff}}
=
n_{\mathrm{g}}
\frac{\Delta\lambda_{\mathrm{th}}}{\lambda_0}.
\label{eq:index_change_from_transient_shift}
\end{equation}

Taking \(n_{\mathrm{g}}=2.4\) and a fitted resonance wavelength \(\lambda_0 = 1.08~\mu\mathrm{m}\) gives
\begin{equation}
\boxed{
\Delta n_{\mathrm{eff}}
\simeq
1.4\times10^{-6}
},
\label{eq:slow_effective_index_change}
\end{equation}

Similarly, for the slow, free-carrier mediated blue shift \(\lvert A_{\mathrm{slow}}\rvert=0.5~\mathrm{mV}\), we get a 
resonance shift \(\Delta\lambda_{\mathrm{fc}}\simeq -0.22~\mathrm{pm}\), corresponding to an absolute change of effective refractive index \(\Delta n_{\mathrm{eff}} \simeq 0.5\times10^{-6} \).

%We note that in Fig.~\ref{fig:transient_response}(a) of the main text, the slow shift has larger magnitude, very close to the fast, thermal shift. While quantitatively different, this remains within in the same order of magnitude as the values above.

\subsection{Thermo-optic estimate}
\label{sec:thermal_absorption_estimate}

We now focus on the thermally induced spectral shift. 
It contains contributions from the temperature dependence of the refractive index and the thermal expansion of the suspended structure,
\begin{equation}
\frac{\left|\Delta\lambda_{\mathrm{th}}\right|}{\lambda_0}
\simeq
\left(
\frac{1}{n}\frac{\mathrm{d}n}{\mathrm{d}T}
+\alpha_L
\right)
\Delta T_{\mathrm{th}},
\label{eq:thermal_shift_with_expansion}
\end{equation}
where \(n=2.4\) and \(\alpha_L\) is the linear thermal-expansion coefficient. Using \(\mathrm{d}n/\mathrm{d}T=1.0\times10^{-5}~\mathrm{K^{-1}}\) and \(\alpha_L=1.0\times10^{-6}~\mathrm{K^{-1}}\) at room temperature~\cite{Stoupin2011ThermalExpansion}, together with \(\lambda_0 = 1.08~\mu\mathrm{m}\), gives an average temperature increase
\begin{equation}
\boxed{
\Delta T_{\mathrm{th}}
\simeq
0.11~\mathrm{K}
}.
\label{eq:transient_temperature_rise}
\end{equation}

For an order-of-magnitude estimate of the required heating power, the suspended nanobeam is treated as a one-dimensional diamond rod of cross-sectional area \(S=wt\), with width $w$ and thickness $t$, connected to a heat sink at each end. For a heating source localized at its center, the temperature profile is linear on each side of the heating spot. The maximum temperature increase at the center $\Delta T_{\max}$ is given by the relation
\begin{equation}
P_{\mathrm{heat}}
=
\frac{4S\kappa_{\mathrm{eff}}}{L}
\Delta T_{\max}.
\label{eq:one_dimensional_heating_power}
\end{equation}
where $P_{\mathrm{heat}}$ is the applied heating power. 
The average temperature $\Delta T_{\mathrm{th}}$ depends on the exact effective length of the cavity. In our geometry, the suspended nanobeam length is $L\simeq60$~\textmu m and the spacer only occupies a central 20~\textmu m-long portion. Thus, we can assume that \(\Delta T_{\mathrm{avg}}=(5/6)\Delta T_{\max}\).
Using \(w=437~\mathrm{nm}\), \(t=340~\mathrm{nm}\) and an effective thermal conductivity \(\kappa_{\mathrm{eff}}=1200~\mathrm{W\,m^{-1}K^{-1}}\) due to the sub-micron cross-section dimensions~\cite{goblot2026heatTransportDiamond} %, and the assumed relation \(\Delta T_{\mathrm{th}}=(5/6)\Delta T_{\max}\) 
gives a heating power
\begin{equation}
P_{\mathrm{heat}}
\simeq
1.58~\mu\mathrm{W}.
\label{eq:circular_heating_power}
\end{equation}
This corresponds to \(P_{\mathrm{heat}}/P_{\mathrm{in}}\simeq0.15\%\) of the total incident green power. The circular beam has a geometric nanobeam overlap of \(f_{\mathrm{ov}}\simeq0.730\) (see section~\ref{sec:green_excitation_profile}), giving
\begin{equation}
\frac{P_{\mathrm{heat}}}{f_{\mathrm{ov}}P_{\mathrm{in}}}
\simeq
0.002
\label{eq:overlap_normalised_heating_fraction}
\end{equation}
or an absorption fraction for the power incident on the waveguide of approximately \(0.2\%\). This estimate is sensitive to the assumed thermal conductivity, one-dimensional heat-flow geometry, contact resistance, excitation power calibration and spectral shift conversion calibration, and is therefore used only as an order-of-magnitude consistency check.

Note that, in the case of elliptical beam geometry as used in Fig.~\ref{fig:transient_response}(a) of the main text, the exact temperature profile in the suspended nanobeam might differ, but the average temperature increase is expected to be on the same order of magnitude for the same incident green power.

For comparison, the reported \(\mathrm{NV}^{-}\) absorption cross-section at \(532~\mathrm{nm}\) is \(\sigma_{532}=3.1\times10^{-17}~\mathrm{cm^2}\)~\cite{Wee2007}. The nominal \(4~\mathrm{ppm}\) NV concentration corresponds to an NV number density
\begin{equation}
n_{\mathrm{NV}}
=
4\times10^{-6}n_{\mathrm{C}}
=
4\times10^{-6}
\left(1.76\times10^{23}~\mathrm{cm^{-3}}\right)
\simeq
7.0\times10^{17}~\mathrm{cm^{-3}},
\label{eq:four_ppm_nv_density}
\end{equation}
where \(n_{\mathrm{C}}\) is the atomic number density of diamond. Assuming, as an upper bound, that the complete nominal NV population is in the negative charge state, the power absorbed by the \(\mathrm{NV}^{-}\) ensemble is
\begin{equation}
P_{\mathrm{abs,NV}}
\lesssim
n_{\mathrm{NV}}\sigma_{532}t
f_{\mathrm{ov}}P_{\mathrm{in}}
\simeq
0.59~\mu\mathrm{W}.
\label{eq:nv_absorbed_power}
\end{equation}
This corresponds to approximately \(0.054\%\) of the total incident power. The inferred heating power responsible for the thermal shift is therefore approximately \(2.7\) times larger than the upper-bound NV-absorbed power. Within this simplified model, direct absorption by the NV ensemble alone is insufficient to explain the inferred heating, suggesting that additional absorption channels associated with substitutional nitrogen or other defects may contribute~\cite{Barry2024SensitiveMagnetometry,Heremans2009PhotoexcitedElectrons}.

\subsection{Drude-equivalent electron-density estimate}
\label{sec:SI_transient_carrier_density}

Finally, we consider the free-carrier plasma dispersion.
The Drude model, accounting only for free electrons for simplicity (free holes also contribute, in general), gives the relation between the change in effective refractive index and the cavity-averaged change in free-electron density $\Delta \tilde N_e$:
\begin{equation}
\Delta n_{\mathrm{eff}}
=
-
\frac{e^2\lambda_0^2}
{8\pi^2c^2\varepsilon_0 n m_{\mathrm{opt}}^*}
\Delta \tilde N_e.
\label{eq:electron_drude_index_change}
\end{equation}
with $e$ the elementary charge, $\varepsilon_0$ the vacuum permittivity and $m_{\mathrm{opt}}^*$ the optical effective mass.
From the measured transverse and longitudinal electron effective masses in diamond, \(m_t=0.2804 \ m_0\) and \(m_l=1.560 \ m_0\)~\cite{Naka2013CarrierMass}, where $m_0$ is the free-electron mass, the optical effective mass is
\begin{equation}
m_{\mathrm{opt}}^*
=
\frac{3}{2/m_t+1/m_l}
=
0.386 \ m_0.
\label{eq:diamond_optical_effective_mass}
\end{equation}
Using \(n=2.4\), \(\lambda_0 = 1.08~\mu\mathrm{m}\) and the experimental value \(\Delta n_{\mathrm{eff}} \simeq 0.5\times10^{-6} \) obtained above, gives
\begin{equation}
\boxed{
\Delta \tilde N_e
\simeq
8.7\times10^{14}~\mathrm{cm^{-3}}
}.
\label{eq:transient_equivalent_carrier_density}
\end{equation}
% The factor-of-two spectral-shift interval in Eq.~\eqref{eq:estimated_carrier_shift_range} corresponds to
% \begin{equation}
% 1.2\times10^{15}~\mathrm{cm^{-3}}
% \lesssim
% \Delta N_e
% \lesssim
% 4.6\times10^{15}~\mathrm{cm^{-3}}.
% \label{eq:carrier_density_range}
% \end{equation}
% The central estimate is approximately \(0.33\%\) of the nominal \(4~\mathrm{ppm}\) NV density, with the factor-of-two interval corresponding to approximately \(0.17\%\)--\(0.66\%\). 
The order of magnitude of $10^{15}~\mathrm{cm^{-3}}$ free electrons is well below the nominal NV density of $7.0\times10^{17}~\mathrm{cm^{-3}}$, and is therefore compatible with ionization or redistribution of a small fraction of the available defects. The above conclusion holds even at the highest excitation power that we could probe, \(P_{\mathrm{in}}\simeq11~\mathrm{mW}\), where we estimate $\Delta \tilde N_e \sim 10^{16}~\mathrm{cm^{-3}}$.
%This comparison establishes plausibility but does not identify which defect species supplies the carriers.

\subsection{Estimation of free-carrier mobility}
\label{subsec:matched_mobility_estimate}

We can take the analysis one step further and extract an estimate for the equivalent effective electron mobility \(\mu_{e,\mathrm{eff}}\).
Within the electron-only Drude model, the estimated index change and absorption change are attributed to the same electron population. Taking their ratio eliminates the carrier density and gives the effective electron mobility:
\begin{equation}
\mu_{e,\mathrm{eff}}
=
\frac{
2e\left|\Delta n_{\mathrm{eff}}\right|
}{
c\,m_{\mathrm{opt}}^*\Delta\alpha_{\mathrm{eff}}
}.
\label{eq:matched_effective_mobility}
\end{equation}
Assuming that the green-induced change in internal loss arises entirely from additional free-carrier absorption, the cavity-averaged absorption change \( \Delta\alpha_{\mathrm{eff}} \) is directly given by:
\begin{equation}
\Delta\alpha_{\mathrm{eff}}
=
\frac{n_{\mathrm{g}}}{c} 2\pi \Delta\gamma_i.
\label{eq:loss_rate_to_absorption}
\end{equation}
The added loss \( \Delta\gamma_i\) has been extracted from the fits of the LIA traces, as detailed in the main text. For the elliptical beam configuration in Fig.~\ref{fig:modulation_depth}b, with \( P_{\mathrm{in}} = 3.28~\mathrm{mW} \), we have
\begin{equation}
\Delta\nu_i
=
3.2~\mathrm{GHz}.
\label{eq:lockin_loss_at_reference_power}
\end{equation}
For \(n_{\mathrm{g}}=2.40\), Eq.~\eqref{eq:loss_rate_to_absorption} gives
\begin{equation}
\Delta\alpha_{\mathrm{eff}}
\simeq
1.6~\mathrm{cm^{-1}}.
\label{eq:lockin_absorption_change}
\end{equation}

The change in refractive index for the same experimental configuration can be extracted from the time trace of Fig.~\ref{fig:transient_response} a, using the procedure described above (section~\ref{subsec:transient_index_change}). We evaluate the magnitude of the blue shift after 25~\textmu s of green excitation, corresponding to the modulation speed of 20~kHz from Fig.~\ref{fig:modulation_depth}b. We find:
\begin{equation}
\Delta n_{\mathrm{eff}}
\simeq 3 \times 10^{-5}
.
\label{eq:elliptical_n_eff}
\end{equation}

The effective mobility required for the same electron population to account for both the estimated index change and the fitted cavity broadening is therefore
\begin{equation}
\boxed{
\mu_{e,\mathrm{eff}}
\simeq
5.7~\mathrm{cm^2\,V^{-1}\,s^{-1}}
}.
\label{eq:effective_mobility_result}
\end{equation}

The analysis above assumes an electron-only Drude response and assigns the fitted internal-linewidth change entirely to absorption. Trapping, space-charge fields, defect-assisted absorption, and differences between the transient and LIA dynamics can all modify the inferred value. The extracted carrier density and mobility should therefore be interpreted as Drude-equivalent consistency estimates rather than intrinsic material parameters.

\section{Estimate of singlet-induced absorption and cavity-linewidth broadening}
\label{sec:singlet_absorption_model}

To estimate the contribution of NV$^-$ singlet absorption to the green-induced cavity loss, we calculated the singlet populations under $532~\mathrm{nm}$ excitation without microwave driving. The calculation uses an eight-level rate-equation model comprising two ground and two excited triplet spin manifolds of NV$^-$, the upper and lower singlet states, and the ground and excited states of NV$^0$. The NV$^0$ levels are included to account for photoionization of NV$^-$ and recombination back to NV$^-$, which changes the fraction of centres available to populate the NV$^-$ singlet states. At each local green intensity, the steady-state populations $p_i$ are obtained by setting $\dot p_i=0$, with $\sum_{i=1}^{8}p_i=1$. The eight-level rate-equation model adopted here is described in detail in the Supporting Information of Kim et al.~\cite{Kim2021Metasurface}.

The local green intensity and NV$^-$ excitation rate are
\begin{equation}
    I_g(z)=\frac{2P_{\mathrm{in}}}{\pi w_x w_z}
    \exp\!\left(-\frac{2z^2}{w_z^2}\right),
    \qquad
    W_g(z)=\frac{\sigma_g I_g(z)}{hc/\lambda_g},
\end{equation}
where $P_{\mathrm{in}}$ is the green-on power after the objective, $\lambda_g=532~\mathrm{nm}$, and $\sigma_g=3.1\times10^{-21}~\mathrm{m^2}$~\cite{Wee2007}. The Gaussian $1/e^2$ intensity radii are $w_x=w_z=0.381~\mu\mathrm{m}$ for circular illumination and $(w_x,w_z)=(0.381,8.923)~\mu\mathrm{m}$ for elliptical illumination. Here, $z$ follows the nanobeam axis. %The excitation is approximated as uniform across the nanobeam cross-section, using the intensity at the transverse beam centre.
Since no microwave field is applied in our experiment, the microwave transition rate is set to zero. Green excitation populates the excited triplet states, which subsequently feed the singlet metastable state through intersystem crossing. Labelling the excited triplet populations $p_3$ and $p_4$, and the upper and lower singlet populations $p_5$ and $p_6$, respectively, the weak-probe steady-state relations are
\begin{equation}
p_6=\frac{k_{35}p_3+k_{45}p_4}{k_{61}+k_{62}},
\qquad
p_5=\frac{k_{35}p_3+k_{45}p_4}{\gamma_s}.
\end{equation}
We use the triplet radiative rates $k_{31}=k_{42}=66\times10^6~\mathrm{s^{-1}}$, the intersystem-crossing rates $k_{35}=7.9\times10^6~\mathrm{s^{-1}}$ and $k_{45}=53\times10^6~\mathrm{s^{-1}}$, and the upper-singlet relaxation rate $\gamma_s=10^9~\mathrm{s^{-1}}$ employed in Refs.~\cite{Dumeige2013,Kim2021Metasurface}. The lower-singlet decay rates are rescaled to reproduce the room-temperature lifetime $(k_{61}+k_{62})^{-1}=219~\mathrm{ns}$ measured by Acosta \textit{et al.}~\cite{Acosta2010Singlet}, while retaining the branching ratio $k_{61}:k_{62}=1:0.7$ used in Refs.~\cite{Dumeige2013,Kim2021Metasurface}.

For the NV$^0$ states, the green excitation cross-section $\sigma_{\mathrm{NV^0}}=6\times10^{-21}~\mathrm{m^2}$ and radiative decay rate $\Gamma_{\mathrm{NV^0}}=53\times10^6~\mathrm{s^{-1}}$ are taken from Kim \textit{et al.}~\cite{Kim2021Metasurface}. In the present calculation, the two ionization rates from the NV$^-$ excited states and the two recombination rates from excited NV$^0$ to the NV$^-$ ground states are each fixed at $10^6~\mathrm{s^{-1}}$. Finally, the infrared intensity is set to zero in the population equations, corresponding to the weak-probe limit in which the probe does not modify the steady-state populations.

Net infrared absorption is proportional to $p_6-p_5$, accounting for absorption from the lower singlet and stimulated emission from the upper singlet. To include the green-beam geometry, we first calculate the local Gaussian intensity $I_g(z)$ along the nanobeam and solve the steady-state rate equations at each position to obtain $p_6(z)$ and $p_5(z)$. The cavity-averaged net singlet fraction is then
\begin{equation}
    \overline{s}(P_{\mathrm{in}})
    =\frac{1}{L_{\mathrm{eff}}}
    \int_{-L_{\mathrm{cav}}/2}^{L_{\mathrm{cav}}/2}
    \left[p_6(z)-p_5(z)\right]\,\mathrm{d}z,
\end{equation}
where $L_{\mathrm{cav}}=20~\mu\mathrm{m}$ is the physical cavity length and $L_{\mathrm{eff}}=32.6~\mu\mathrm{m}$ includes the cavity energy extending into the Bragg mirrors.

The integration covers the full physical cavity, while the finite extent of the green illumination enters through the position-dependent populations. For elliptical illumination, integrating the normalized Gaussian intensity over this region gives an effective illuminated length of approximately $10.9~\mu\mathrm{m}$. We assume uniform populations across the nanobeam cross-section and unity transverse optical confinement.

The wavelength-dependent singlet absorption cross-section is approximated by
\begin{equation}
    \sigma_{\mathrm{IR}}(\lambda)
    =\frac{\sigma_{\mathrm{IR},0}}
    {1+4\left[(\lambda-\lambda_s)/\Delta\lambda_s\right]^2},
\end{equation}
with $\sigma_{\mathrm{IR},0}=3\times10^{-22}~\mathrm{m^2}$, $\lambda_s=1042~\mathrm{nm}$, and a full width at half maximum $\Delta\lambda_s=11~\mathrm{nm}$~\cite{Probst2026SingletTransition}. 

Taking a total NV density, including both charge states, of $N_{\mathrm{NV}}=7.04\times10^{17}~\mathrm{cm^{-3}}$ corresponding to $4~\mathrm{ppm}$, the effective power-absorption coefficient and added internal linewidth are
\begin{equation}
    \Delta\alpha_{\mathrm{eff}}
    =N_{\mathrm{NV}}\sigma_{\mathrm{IR}}(\lambda)\overline{s},
    \qquad
    \Delta\gamma_i
    =\frac{\Delta\kappa_i}{2\pi}
    =\frac{c}{2\pi n_g}\Delta\alpha_{\mathrm{eff}},
\end{equation}
where $n_g=2.40$, $\Delta\kappa_i$ is the added cavity-energy decay rate, and $\Delta\gamma_i$ is its contribution to the full linewidth in ordinary-frequency units. Since the singlet populations vanish in the steady-state green-off limit, this gives the predicted green-on minus green-off linewidth change due to singlet absorption alone.

\begin{figure}[htbp]
    \centering
    \includegraphics[width=0.9\textwidth]{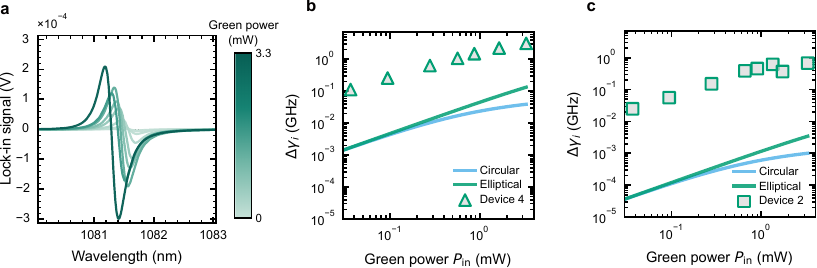}
    \caption{
    \textbf{Comparison of measured green-induced cavity loss with predicted NV$^-$ singlet absorption.}
    \textbf{a}, Wavelength-resolved lock-in spectra of Device~2 near $1081~\mathrm{nm}$ at increasing green excitation powers. The signal is the signed, phase-rotated lock-in quadrature $X_{\mathrm{rot}}$; the colour scale indicates the green-on power after the objective.
    \textbf{b,c}, Added internal cavity linewidth $\Delta\gamma_i=\Delta\kappa_i/(2\pi)$ as a function of incident green power $P_{\mathrm{in}}$ for Device~4 near $1045~\mathrm{nm}$ (\textbf{b}, open triangles) and Device~2 near $1081~\mathrm{nm}$ (\textbf{c}, open squares). Symbols denote values extracted from fits to the lock-in spectra. Solid curves show the singlet-absorption predictions for circular (blue) and elliptical (green) illumination, calculated using the eight-level steady-state model without microwave driving, a total NV density of $4~\mathrm{ppm}$, and the weak-probe approximation.
    }
    \label{fig:singlet_absorption_comparison}
\end{figure}

Figure~\ref{fig:singlet_absorption_comparison} compares the predicted singlet-induced linewidth broadening with the internal-loss changes extracted from the wavelength-resolved lock-in measurements. Panel~\textbf{a} shows the measured spectra near $1081~\mathrm{nm}$ for Device~2, where an asymmetric response remains visible despite the large detuning from the singlet transition. Panels~\textbf{b} and \textbf{c} compare the extracted $\Delta\gamma_i$ for Device~4 near $1045~\mathrm{nm}$ and Device~2 near $1081~\mathrm{nm}$, respectively, with the predictions for circular and elliptical green illumination. At low excitation powers, the two illumination geometries give similar predicted broadening. At higher powers, elliptical illumination produces a larger cavity-averaged singlet population because it distributes the excitation over a longer section of the nanobeam, reducing local saturation. The predicted broadening is substantially smaller near $1081~\mathrm{nm}$ owing to the reduced singlet absorption cross-section away from the transition frequency.
For both devices, the extracted internal-loss changes substantially exceed the calculated singlet contribution. Thus, although singlet absorption may contribute to the larger loss observed closer to the infrared transition, it cannot account for the magnitude of the measured loss within the assumptions of this model. 

Two other observations support that an additional free-carrier mechanism is required to explain the observations. First, the population in the singlet state is expected to relax within less than 1~$\mu$s after the green excitation is switched off, which is much faster than the transient measured in Fig.~\ref{fig:transient_response}c. Second, it is hard to see why populating the singlet should cause a blue shift of the resonance. In the conventional Lorentz oscillator model, the refractive index increases at energies just below a resonance (wavelength $\geq 1042$~nm), but we find that it decreases (blue shift of the resonance).

\end{document}